\documentclass[onecolumn,preprintnumbers,amsmath,amssymb]{revtex4}
\usepackage{amsmath}
\usepackage{extarrows}
\usepackage[ruled]{algorithm2e}
\usepackage{graphicx}
\usepackage{dcolumn}
\usepackage{bm}
\usepackage[all]{xy}
\usepackage{indentfirst}
\usepackage{amsmath}
\usepackage{multirow}
\usepackage{mathrsfs}
\usepackage{bbm}
\usepackage{euscript}
\usepackage{amssymb}
\usepackage{extarrows}
\usepackage{bbm}
\usepackage[ruled]{algorithm2e}
\usepackage{graphicx}
\usepackage{dcolumn}
\usepackage{bm}
\usepackage[all]{xy}
\usepackage{indentfirst}
\usepackage{amsmath}
\usepackage{multirow}
\usepackage{mathrsfs}
\usepackage{euscript}
\usepackage{amssymb}
\begin{document}
\linespread{1.2}
\title{A nonlocal game for witnessing quantum networks}

\author{Ming-Xing Luo}

\affiliation{\small{} $^1$The School of Information Science and Technology, Southwest Jiaotong University, Chengdu 610031, China;
\\
\small{} $^2$CSNMT, International Cooperation Research Center of China, Chengdu 610031, China}

\begin{abstract}
Nonlocal game as a witness of the nonlocality of entanglement is of fundamental importance in various fields. The red well-known nonlocal games or equivalent linear Bell inequalities are only useful for Bell networks consisting of single entanglement. Our goal in this paper is to propose a unified method for constructing cooperating games in network scenarios. We  propose an efficient method to construct multipartite nonlocal games from any graphs. The main idea is the graph representation of entanglement-based quantum networks. We further specify these graphic games with quantum advantages by providing a simple sufficient and necessary condition. The graphic games imply a linear Bell testing of the nonlocality of general quantum networks consisting of EPR states. It also allows generating new instances going beyond CHSH game. These results have interesting applications in quantum networks, Bell theory, computational complexity, and theoretical computer science.
\end{abstract}
\maketitle

\section*{Introduction}

A nonlocal game is generally described as multiple space-separated players interacting with a referee. The communication between players are forbidden. Remarkably, it is related to Bell theory that is fundament in quantum mechanics ${}^{1}$ when players are allowed to share classical resources or quantum entangled resources. Clauser-Horne-Shimony-Holt (CHSH) game as a distributed evaluating of Boolean equation: $y_1\oplus y_2=x_1\wedge{}x_2$, provides the first notable way for witnessing the nonlocality of   Einstein-Podolsky-Rosen (EPR) state, $^{1,2,3,4,5,6}$ where $x_i, y_i$ are respective binary input and output. The shared entanglement permits a quantum strategy with larger winning probability than all classical strategies.

Generally, a nonlocal game allows all the players to determine a joint strategy from the complete knowledge of inputs distribution and the predicate conditions for win. The corresponding procedure can be featured by a generalized hidden variable model with classical sources or quantum sources.$^{1}$ One fundamental problem is to determine whether quantum mechanics is superior to classical theory in terms of nonlocal tasks. These evaluations are equivalent to solving special optimization problems from Tsirelson's reductions. $^{7}$ Unfortunately, no efficient algorithm exists for general multipartite nonlocal games. $^{8,9,10,11,12,13}$ Besides verifying entanglement, nonlocal games have so far inspired numerous applications, such as multi-prover interactive proofs, $^{14,15,16}$ quantum proof verification, $^{17,18}$ hardness of approximation, $^{19,20,21}$ PCP conjecture, $^{22,23,24}$ separating correlations, $^{25,26,27,28}$ and communication complexity theory.$^{29,30}$

There are various interesting nonlocal games except for CHSH games. XOR games are the most well studied.$^{31,32,33,34}$ The global task is to XOR of the outcomes of all players. Other games include Kochen-Specker game,$^{4,31}$ non-zero sum games,$^{35,36}$ magic square game,$^{37}$ graph isomorphism game,$^{38}$ Bayesian game,$^{39}$ and conflicting interest game.$^{40}$ The key to achieve a quantum advantage is the nonlocal correlations generated by local measurements on the shared entangled states. Similar correlations are not achievable by remote players using local strategies with any shared classical resources. These tasks show the important role of quantum entanglement in information processing.$^{41}$ Comparison to single entanglement,$^{1}$ the network scenarios of multiple independent sources require the nonlinear Bell inequalities to test the nonlocality.$^{42,43,44,45}$ Hence, how to construct meaningful nonlocal games should be interesting in scaling applications of entangled resources. Two related problems are as follows:

{\bf Problem 1}. How to efficiently construct nonlocal games for space-separated players who share a quantum network consisting of entangled states?

{\bf Problem 2}. How to characterize nonlocal games with or without quantum advantages in network scenarios?

The goal of this paper is to address these problems in the context of cooperative nonlocal games, as shown in Fig.1. We propose a unified method to construct multipartite games from graph representations of general multi-source networks. Each game is determined by some subgraphs that can be constructed efficiently. Note that for any graph with $N$ nodes, the proposed method provides $O(4^{nN})$ different nonlocal games with $n$ players. These graphic nonlocal games will further be classified in terms of the quantum advantage using a simple sufficient and necessary condition. The result holds for the same probability distribution of all inputs going beyond previous algorithmic results $^{47}$ or semi-definite programs.$^{7,10,11,48,49}$ Surprisingly, the present graphic games allow testing the nonlocality of multi-source quantum networks going beyond previous CHSH game for single entanglement.$^{1,2,3,36}$ Compared with the recent nonlinear witness,$^{42,43,44,45}$ the graphic game provides the first verification of general networks using linear Bell testing beyond semiquantum game.$^{46}$ The new result is also useful for nonlocal satisfiability problems going beyond CHSH game $^{3}$ and cubic game.$^{37}$ The graphic game finally extends the guessing your neighbor's input (GYNI) game.$^{50,51}$ The present model provides novel instances to feature nonlocal games with different performances.$^{52}$

\begin{figure}
\begin{center}
\resizebox{200pt}{155pt}{\includegraphics{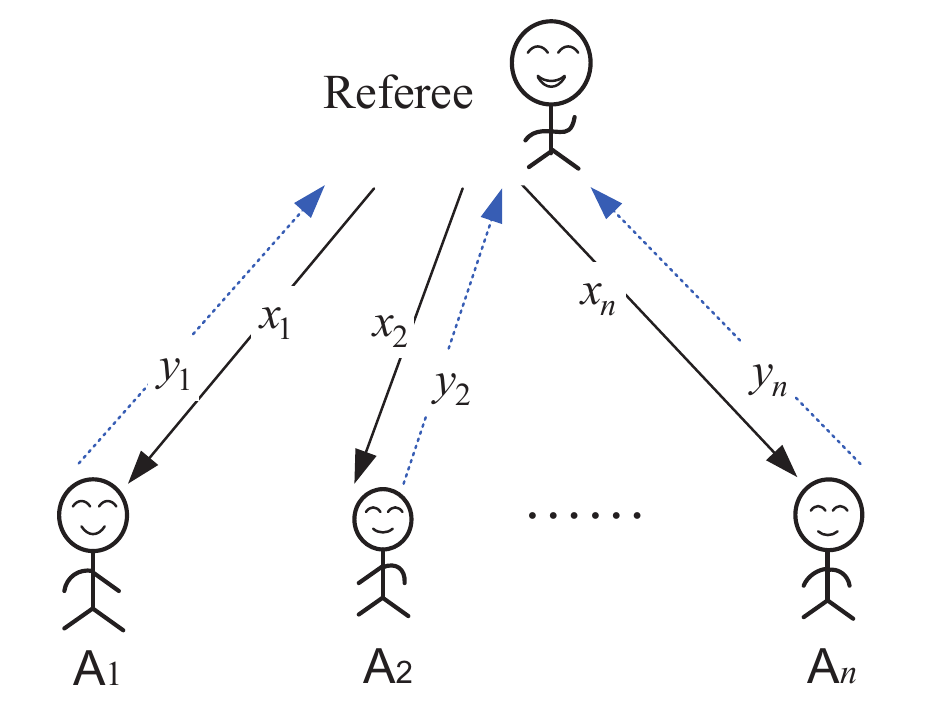}}
\end{center}
\caption{\small{}Cooperating nonlocal game. A referee chooses questions $x_k$s from a finite set according to a prior distribution $p(\textbf{x})$, and sends to players. Each player $\textsf{A}_k$ should output an answer $y_k$ based on the received question $x_k$ and the shared resource without communicating with other players. The referee then evaluates a payoff function $F$ with all outputs and inputs to determine win or lose for all the players.}
\end{figure}

\section*{Result}

\subsection*{Nonlocal multipartite cooperating games}

An $n$-player nonlocal cooperating game consists of $n$ players $\textsf{A}_1, \textsf{A}_2, \cdots, \textsf{A}_n$ and a referee,$^{31}$ as shown in Fig.1. All the players agree with a joint strategy beforehand but cannot communicate
with each other during the game.  The referee firstly chooses $n$ questions: $x_1, x_2, \cdots, x_n$ from a finite set ${\cal X}:=X_1\times X_2\times \cdots \times X_n$ according to a known distribution $p(\textbf{x})$. And then,  sends $x_i$ to $\textsf{A}_i$, $i=1, 2, \cdots, n$. All the players are now required to reply with answers $y_1, y_2, \cdots, y_n$ chosen from a finite set ${\cal Y}:=Y_1\times Y_2\times \cdots \times Y_n$ to the referee. Finally, the referee determines whether all the players win or lose the game according to a payoff function: $F: {\cal X}\times {\cal Y}\to \{0, 1\}$, i.e., win if $F(\textbf{x}, \textbf{y})=1$ and lose if $F(\textbf{x},\textbf{y})=0$, where $\textbf{x}=x_1x_2\cdots x_n$, $\textbf{y}=y_1y_2\cdots y_n$. The optimal average winning probability of all the players, i.e., the nonlocal value of the game, is defined by:

\begin{eqnarray}
\varpi_c=\sup_{\Omega}
\sum_{\textbf{x},\textbf{y},\lambda}
p(\textbf{x})\mu(\lambda)
F(\textbf{x},\textbf{y})
P(\textbf{y}|\textbf{x},\lambda)
\label{eqn-1}
\end{eqnarray}
Here, $\mu(\lambda)$ is the probability measure of the hidden classical resource $\lambda$ and satisfies $\sum_{\lambda}\mu(\lambda)=1$. $P(\textbf{y}|\textbf{x},\lambda)$ denotes the joint conditional probability depending on the shared randomness $\lambda$. The supremum is over all the possible classical spaces $\Omega$ of hidden variables $\lambda$. Generally, it is hard to approximate the nonlocal value $\varpi_c$ of multipartite games.$^{10,11,49}$ From the linearity of $\varpi_c$ with respect to all probability distributions, it is sufficient use deterministic strategies for all the players, i.e., $p(\lambda)=1$ for some $\lambda$.

In quantum scenarios, players are
allowed to share various entangled
states and perform quantum measurements (with noncommuting operators which are different from commuting operators in classical physics) to obtain answers. Let $\rho$ be the shared entangled state. Denote positive-operator valued measurements (POVMs) of all the observers as $\{M_{y_1}^{x_1}\}$, $\{M_{y_1}^{x_2}\}$, $\cdots, \{M_{y_n}^{x_n}\}$, which depend on the received questions $x_1, x_2, \cdots, x_n$. These operators satisfy the normalization conditions: $\sum_{y_i}M_{y_i}^{x_i}=\mathbb{I}$ for $i=1, 2, \cdots, n$, where $\mathbb{I}$ is the identity operator. The optimal winning probability for all the quantum players to player a cooperating nonlocal game defined above is defined by:
\begin{eqnarray}
\varpi_q=\sup
\sum_{\textbf{x},\textbf{y}}
p(\textbf{x})F(\textbf{x},\textbf{y})
{\rm{}Tr}[\otimes_{j=1}^nM_{y_j}^{x_j}
\rho],
\label{eqn-2}
\end{eqnarray}
where the supremum is over all the possible quantum states and local POVMs. Note that each linear Bell-type inequality is equivalent to a cooperating nonlocal game.$^{53}$ The nonlocal value $\varpi_q$ is related to the Tsirelson bound of linear Bell inequality.$^{7}$ A central problem in the non-locality theory is to find computable good bounds of nonlocal value $\varpi_q$.$^{7,31}$ Unfortunately, it is QMA-hard to approximate the nonlocal value of $\varpi_q$.$^{17,53}$

\subsection*{Graphic games}

CHSH game provides a novel idea to witness a bipartite entanglement.$^{2,3}$ As a natural extension, it would be of interesting to characterize multiple entangled states shared by space-separated observers in a network scenarios. Although the linear testing of single entanglement$^{53,55,56}$ allows designing equivalent nonlocal game, it is unknown how to construct meaningful nonlocal games for multi-source quantum networks that requires nonlinear testing $^{42,43,44,45}$ or semiquantum testing.$^{46}$ To address this problem, we propose a general method to construct nonlocal games from any graphs with only vertices. Surprisingly, the graphic games can be completely classified with respect to the nonlocal values. To explain the main result, we firstly present the following definitions.

{\it Definition 1}. (The graphic representation of quantum networks) Consider a multi-source network ${\cal N}_q$ consisting of multiple independent EPR states shared by different observers. Its graphic representation is defined as a vertex graph ${\cal G}_q$, where each node denotes one entanglement.

\begin{figure}
\begin{center}
\includegraphics{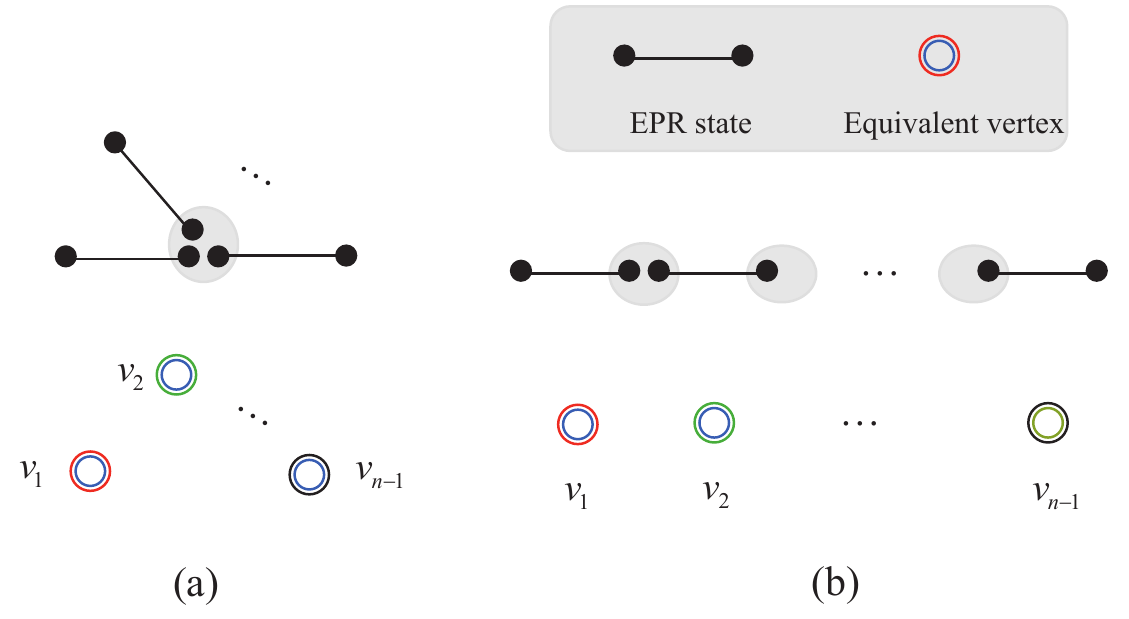}
\end{center}
\caption{\small{}Graphic representation of multi-source quantum networks consisting of EPR states. (a) Star-type network. Each of $n-1$ observers share one EPR state with the center observer. (b) Chain-type network of entanglement swapping in long distance. Any two adjacent observers share one EPR state. Each entanglement is schematically represented by one vertex in graph. The relationship that two observers share one entangled state is equivalent to two players sharing one vertex, where each player owns vertices with the same color.}
\end{figure}

Some typical examples of quantum networks are shown in Fig.2. Generally, for any graph, we can assign a subgraph to each player accord to the input. In this case, the relation that two observers share one entangled state is equivalent to two players sharing one vertex, see Fig.2, where each player owns some vertices with the same color. These nonlocal games can be then characterized by using the consistency conditions on the common vertices shared by different players. Formally, the graphic game is defined as follows:

{\it Definition 2}. An $n$-partite graphic game is a six-tuple: $({\cal G}, {\cal A}, {\cal V}, {\cal X}, {\cal Y}, F)$. ${\cal G}$ is a general graph with vertex set $V$  (the edges are not required in this application). ${\cal A}$ denotes the set of all the players $\textsf{A}_i$s and referee.  ${\cal V}$ denotes the set of all the vertex sets $V^{x_i}\subseteq V$ owned by players, where $V^{x_i}$ denotes the set of vertices assigned to the player $\textsf{A}_i$ according to the input question $x_i$. Assume that $V^{x_i}$s satisfy the following conditions: $V^{x_i}\cap V^{x_j}=\emptyset$ for $x_i=x_j=1$ with $1\leq i<j \leq m$, where $m$ is a parameter satisfying $1\leq m<n$. ${\cal X}$ denotes the set of binary problems $x_1, x_2, \cdots, x_n\in \{0,1\}$. ${\cal Y}$ denotes the set of all the assignments on vertices (as outputs) by players. $F: {\cal X}\times {\cal Y}\to \{0, 1\}$ is a payoff function depending on all the inputs and outputs of players. A graphic game is implemented as follows:
\begin{itemize}
\item[]{\bf Input}-The referee randomly chooses binary question $x_i$ according to the distribution $\{p,1-p\}$, and sends to the player $\textsf{A}_{i}$, $i=1, 2, \cdots, n$.
 \item[]{\bf Output}-The player $\textsf{A}_i$ assigns an integer $y_i\in\{\pm 1\}$ on each vertex in the set $V^{x_i}$, and sends all the assignments to the referee secretly.
\item[]{\bf Winning conditions}-All the players win the graphic game, i.e., $F(\textbf{x}, \textbf{y})=1$, if and only if their assignments satisfy the following consistency conditions:
\begin{itemize}
\item[(a)] $S^{x_i}=1$ for $i=m+1, m+2, \cdots, n$;
\item[(b)] $S^{x_i;x_j}=-1$ when $x_i=x_j=1$, or $1$ otherwise, for $1\leq i\leq m<j\leq n$;
\item[(c)] $S^{x_i;x_j}=S^{x_j;x_i}=1$ for $m<i<j\leq n$;
\end{itemize}
 where $S^{x_i}$ denotes the product of all the assignments of the player $\textsf{A}_i$, and  $S^{x_i;x_j}$ denotes the product of all the assignments by the player $\textsf{A}_i$ on the vertices shared with the player $\textsf{A}_j$.
\end{itemize}

Definition 2 contains previous nonlocal games $^{3,31,37}$ as special cases. The consistency conditions (the conditions (b) and (c) in Definition 2) are important for the most of nonlocal games.$^{3}$ Quantum correlations derived from entangled resources can satisfy these kinds of requirements with higher winning probability over all classical achievable correlations. Here, $m$ is a free parameter satisfying $1\leq m<n$, which is used to feature the players whose assignments satisfy the consistency conditions. Specifically, the consistency conditions should be satisfied by the outputs of all the players to win a nonlocal game. Our graphic game is then similar to a nonlocal computation such as CHSH game with nonlinear restrictions shown in conditions (a)-(c). Generally, for a graph ${\cal G}$ with $N$ vertices, there are $2^N-1$ nontrivial subgraphs, which imply $O(4^{nN})$ different graphic games. It is hard to evaluate or approximate the nonlocal values for all these games. Hence, it should be interesting if some restricted games can be distinguished.

\begin{figure}
\begin{center}
\resizebox{250pt}{220pt}{\includegraphics{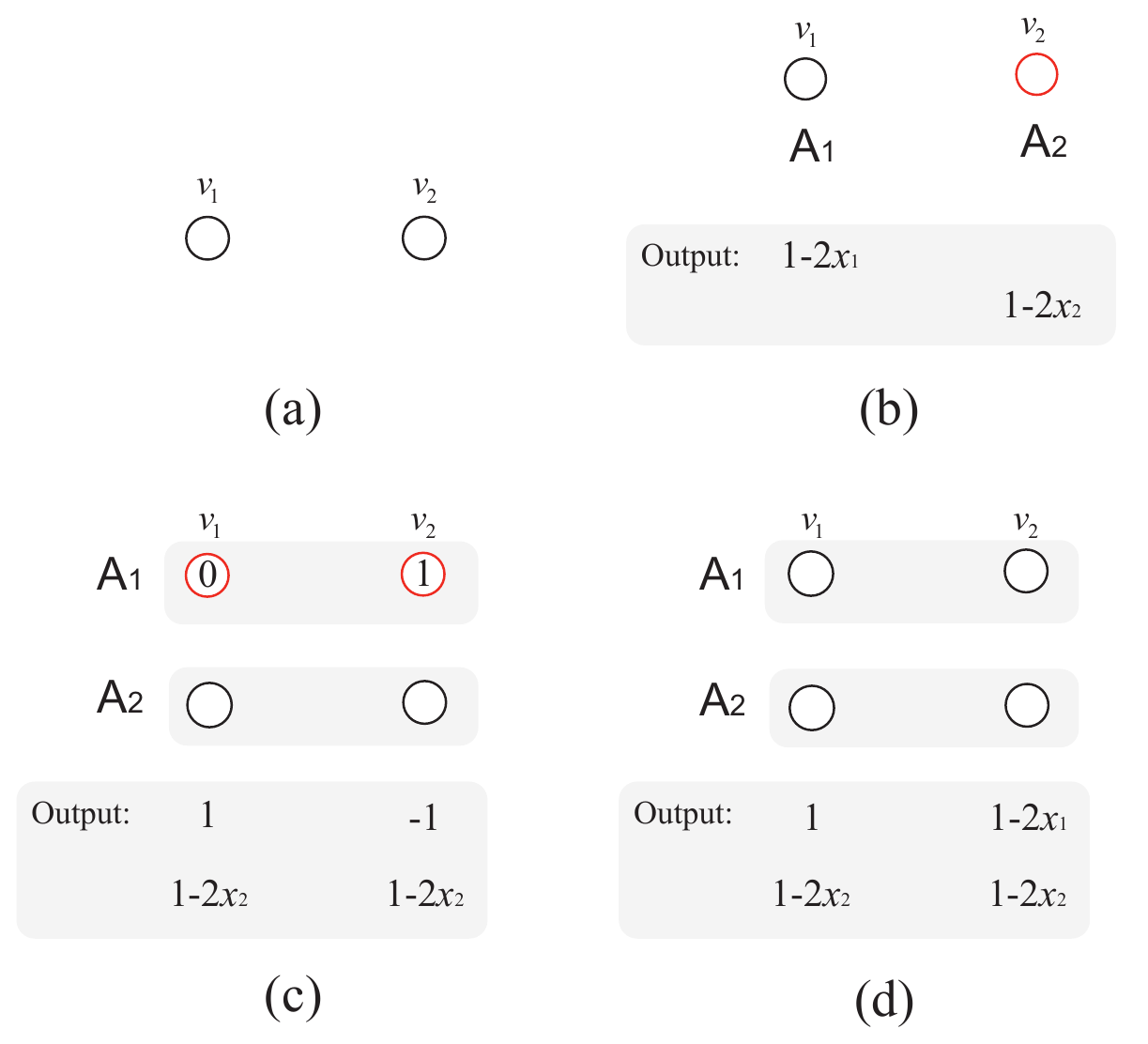}}
\end{center}
\caption{\small{}A graphic nonlocal game. (a) A graph with two vertices. (b) A graphic game with two players. The player $\textsf{A}_i$ holds the vertex $v_i$ independent of the input, $i=1, 2$. (c) A graphic game with two players. The player $\textsf{A}_1$ holds the vertex $v_{x_1}$ depending on its input $x_1$ while the player $\textsf{A}_2$ owns the vertices $v_{1}, v_{2}$ independent of the input. (d) A graphic game with two players. The player $\textsf{A}_i$ holds the vertexes $v_1, v_2$ independent of the input, $i=1, 2$. The players can share classical or quantum resources.}
\end{figure}

\subsection*{Toy example}

Take the graph ${\cal G}$ shown in Fig.3(a) as an example. There are two players in Fig.3(b). Each player holds one vertex independent of inputs, where their outputs are $1-2x_1, 1-2x_2$. There is no consistency requirement for two players with $m=1$. It is easy to obtain $\varpi_c=1$ for classical players. Hence, there is no quantum advantage from $\varpi_q=\varpi_c=1$. For the game in Fig.2(c), the player $\textsf{A}_1$ holds different vertices $v_{x_1+1}$ according to the input $x_1$ while $\textsf{A}_2$ owns the whole graph independent of the input. In this case, two players have to satisfy the consistency requirements in conditions (b) and (c), i.e., $y_{1;A_1}=y_{1;A_2}=1$ and  $y_{2;A_1}=y_{2;A_2}=-1$, where $y_{i;A_j}$ denotes the assignment on the vertex $v_i$ by the player $\textsf{A}_j$. Now, assume that the player $\textsf{A}_1$ assigns $1-2x_1$ on the shared one vertex according to the input $x_1$.  The player $\textsf{A}_2$ assigns the same value of $1-2x_2$ on two vertices. With these assignments, the graphic game is equivalent to CHSH game. $^{3}$  This implies a strict quantum advantage for two players who share an EPR state.$^{2}$ Similar result holds for the graphic game shown in Fig.3(d). To explain the main result, some definitions will be introduced beforehand.

{\it Definition 3.} Consider a graphic game with $n$ players $\textsf{A}_1, \textsf{A}_2, \cdots, \textsf{A}_n$. For each $i$ with $1\leq i\leq m$, ${\cal A}_i$ denotes the set of players $\textsf{A}_{j}$s ($m+1\leq j\leq n$) who share vertices with $\textsf{A}_i$ for each input $x_ix_j$, i.e.,
\begin{eqnarray}
{\cal{}A}_i=\{\textsf{A}_j
|V^{x_i}\cap{V}^{x_j}
\not=\emptyset, \forall x_i, x_j=0, 1\}.
\label{A1}
\end{eqnarray}

{\it Definition 4.} Denote ${\cal{}A}^s_i$ as the set of players as follows:
\begin{eqnarray}
{\cal A}^s_i=\{(\textsf{A}_{i}, \textsf{A}_{i_{1}}, \cdots, \textsf{A}_{i_{s-1}})| \textsf{A}_{i_{j}}\in {\cal A}_i,
 V^{x_i}\cap
(\cap_{j=1}^{s-1}V^{x_{i_j}})
\not=\emptyset, \forall x_i, x_2, \cdots, x_{i_{s-1}}=0, 1\},
\label{A2}
\end{eqnarray}
where each element consists of $s$ players $\textsf{A}_{i}, \textsf{A}_{i_{1}}, \cdots, \textsf{A}_{i_{s-1}}$ who share common vertices for each input $x_ix_{i_1}\cdots{}x_{i_{s-1}}$, and $2\leq s\leq n-m$.

${\cal A}^s_i$ characterizes the consistency requirements among $s$ players of $\textsf{A}_j$s and $\textsf{A}_i$. In what follows, we omit the case that all the players have no common vertex because of  $\varpi_c=\varpi_q$.

{\it Definition 5.} Let $I_i$ be an integer associated with ${\cal{}A}^s_i$ as:
\begin{eqnarray}
I_i=\min \{s |{\cal{}A}^s_i\not=\emptyset\}, \forall i=1, 2, \cdots, m.
\label{A3}
\end{eqnarray}

With these definitions, we find all the graphic games with quantum advantages when proper consistency conditions are satisfied.

{\bf Theorem 1}. For any graphic game with $\varpi_c\not=1$, there is a quantum advantage, i.e., $\varpi_q>\varpi_c$, if and only if $\min\{I_1, I_2, \cdots, I_m\}=2$.

Theorem 1 presents a sufficient and necessary condition for graphic games with quantum advantage. One example is shown in Fig.3(b). This provides the first instance for Problem 2 with a deterministic separation of classical correlations and quantum correlations in correlation space. It is interesting to intuitively show the relationship between $I_i$ and quantum advantage. Note that ${\cal{}A}^s_i$ shown in Eq.(4) features the number of players who have the consistency requirements with $\textsf{A}_i$ independent of inputs. If one divides a network into $m$ subnetworks according to players $\textsf{A}_i$s ($i=1, 2, \cdots, m$) who have no shared vertices, $I_i$ is then used to describe the topology characters of a given network. Specially, it means that the network consists of star-type in Fig.2(a) networks or chain-type subnetworks in Fig.2(b) for $I_i=2$. For $I_i=k>2$, the network consists of various cyclic subnetworks. In this case, similar to previous nonlinear method $^{42,43,44,45}$ Theorem 1 implies that there is no useful linear testing scheme for cyclic networks. One example is triangle network $^{42,45}$. The proof is inspired by the unbalanced CHSH game shown in Supplementary Notes 1 and 2.$^{6,37}$ To explain the main idea, consider a graphic game ${\cal G}$ with $m=1$ consisting of three players $\textsf{A}_1, \textsf{A}_2, \textsf{A}_3$. Two different games ${\cal G}_i$ are defined as follows. ${\cal G}_1$ requires that the outputs of two pairs of players  $\{\textsf{A}_1, \textsf{A}_2\}$, and $\{\textsf{A}_1, \textsf{A}_3\}$ are consistency simultaneously. ${\cal G}_2$ requires that the outputs of three pairs of players $\{\textsf{A}_1, \textsf{A}_2\}$, and $\{\textsf{A}_1, \textsf{A}_3\}$, and $\{\textsf{A}_2, \textsf{A}_3\}$ are consistency simultaneously. Both games can be regarded as two and three simultaneous CHSH games, respectively. Theorem 1 means that ${\cal G}_1$ has a quantum advantage with $I_1=2$ while ${\cal G}_2$ has no quantum advantage with $I_1=3$. The main reason is that there are too many restrictions in ${\cal G}_2$ to achieve a quantum advantage. This provides a new witness for entanglement-based quantum networks$^{45}$ and nonlocal games.$^{33}$

\subsection*{Witnessing multi-source quantum networks}

Multi-source quantum networks can extend applications of single-source Bell network in large scale.$^{60,61}$ However, it is hard to verify these distributive entangled states in global pattern due to the non-convexity of quantum correlations. One useful way is to make use of some nonlinear Bell-type inequalities.$^{42,43,44,45}$ Interestingly, the present model provides another witness for general quantum networks. Specifically, let one vertex schematically represent one EPR state, as shown in Definition 1.$^{2}$ Any multi-source quantum network consisting of EPR states is equivalent to a graph with lots of vertices. Some examples are shown in Fig.2. Fig.2(a) presents a star-type quantum network, where each pair of $\textsf{B}_i$ and $\textsf{A}$ share one EPR state $|\Phi\rangle_i$, $i=1, 2, \cdots, n-1$. Its graphic representation consists of $n-1$ vertices $v_1, v_2, \cdots, v_{n-1}$. The assumption that each pair of players share one EPR state can be schematically represented by sharing one vertex in terms of the graphic game in Definition 2. Similar representation holds for chain-type network shown in Fig.2(b). From the equivalent reformations, there are lots of testing to achieve a quantum advantage when the graphic games are defined. Generally, from Theorem 1 we obtain a corollary as:

{\bf Corollary 1}. For any quantum network ${\cal N}_q$ consisting of EPR states shared by $n$ observers, there are nonlocality witnesses in terms of graphic game if ${\cal N}_q$ is $k$-independent with $k\geq 2$.

In Corollary 1, $k$-independence means that there are a set of $k$ observers who do not share any entanglement among themselves. Note that for verifying quantum networks, all the shared vertices $V^{x_i}$s are independent of inputs. From the equivalent reduction,$^{45}$ any $k$-independent network is equivalent to a generalized star-type network, as shown in Fig.2(a). The proof of Corollary 1 follows easily from the corresponding graphic games, where $\min_iI_i=2$ because no more than two players share vertices. Take the star-type network shown in Fig.2(a) as an example. Each pair of players $\{\textsf{B}_i, \textsf{A}\}$ shares one vertex $v_i$ independent of their inputs, $i=1, 2, \cdots, n-1$. There exists quantum advantage because of $I_1=2$ by assuming $m=1$ (or $m=n-1$). For the chain-type network shown in Fig.2(b), each vertex $v_i$ is owned by two adjacent players $\textsf{A}_i$ and $\textsf{A}_{i+1}$. One can define different $m$ by choosing the players without sharing vertices (independent in quantum networks).$^{45}$ Quantum advantages hold for these graphic games. Generally, the present graphic games allow the first linear testing for multi-source quantum networks consisting of EPR states.

\subsection*{CHSH game}

Two equivalent forms of CHSH game in Ref.[3] are shown in Fig.3(c) and Fig.3(d). Another extension is cube game, which is defined over an $n$-dimensional cube graph ${\cal G}$ with $2^n$ vertices represented by $\{(q_1, q_2, \cdots, q_n), \forall q_i=0, 1\}$, as shown in Fig.4(a), where $m=1$. Specially, the referee randomly chooses binary question $x_i\in \{0,1\}$ according to a given distribution $\{p,1-p\}$, and sends it to the player $\textsf{A}_i$. Each player assigns $1$ or $-1$ to their own vertices on $n-1$-dimensional hyperplane defined by $\{(q_1, q_2, \cdots, q_n)|q_i=x_i\}$  and sends to the referee. All the players win the game if and only if their assignments satisfy the requirements in Definition 2. From Theorem 1, it follows that the cubic game has no quantum advantage over all the classical strategies when $n\geq 3$, where $I_1=n$. Our result is different from a recent result with relaxed consistency conditions.$^{35}$ Another simple extension with $m\geq 1$ is defined as: the player $\textsf{A}_i$ owns the local subgraph defined by $\{(q_1, q_2, \cdots, q_n)|q_1=q_2=\cdots=q_m=x_i\}$ for $i=1, 2, \cdots, m$, and $\{(q_1, q_2, \cdots, q_n)|q_j=x_j\}$ for $j=m+1, m+2, \cdots, n$. Note that any two players of $\textsf{A}_1, \textsf{A}_2, \cdots, \textsf{A}_m$ have no common subgraph for all the inputs. It is straightforward to check that this generalization has quantum advantage when $m=n-1$.

\begin{figure}
\begin{center}
\includegraphics{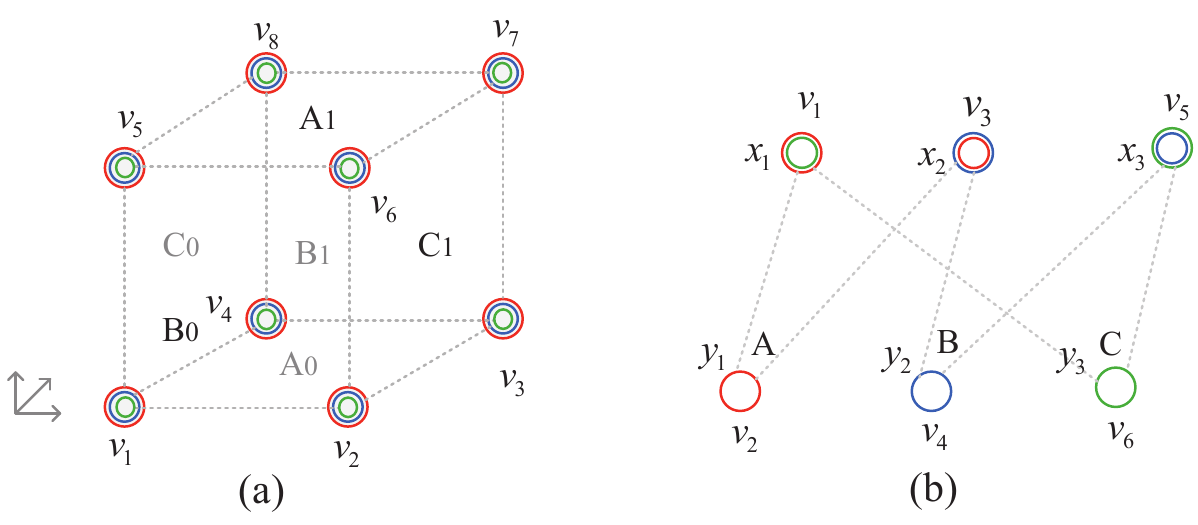}
\end{center}
\caption{\small{}Some graphic games. (a) Cubic game. Each player owns all the vertices on a 2-dimensional plane (denoted by $A_i$, $B_j$, $C_k$) with different colors depending on the input. (b) The equivalent graphic game of GYNI game.$^{50}$ Each player owns three vertices of one subgraph (denoted by $A$, $B$, $C$) with different colors. Here, the edges with dot lines are used for visualization. }
\end{figure}

\subsection*{Distributive SAT problems}

A Boolean satisfiability problem (SAT) aims to check some given equations.$^{62}$ The decision problem is of central importance in theoretical computer science, complexity theory and algorithmic theory.$^{63}$ Interestingly, each graphic game is equivalent to a distributive SAT problem. Take the game shown in Fig.4(a) as an example. Let $y_{i,x_j}$ be assignment on the vertex $v_i$ by $\textsf{A}_j$ according to the input $x_j, i=1, 2, \cdots, 8; j=1, 2, 3$. From Definition 2, the winning conditions are equivalent to Boolean equations with 48 variables. Theorem 1 shows a qualitative decision without quantum advantage from shared entangled resources. Generally, for an $n$-player graphic game involving $N$ vertices, there are $O(nN)$ conditions which should be checked.$^{64}$ Theorem 1 provides an intuitive completion for distributive SAT problems with different resources using graphic game model. Hence, in applications, an equivalent graphic game should be found for special decision problems. One possible way is to reduce the required clauses firstly even if it is hard. And then, define a graphic game clause-by-clause. Special output strategies may be applied as guess your neighbour's input (GYNI) game. $^{50}$

\subsection*{GYNI game}

Guess your neighbour's input (GYNI) game has recently been proposed to separate the multipartite quantum correlations from classical correlations. $^{50}$ This game is equivalently represented by a graphic game, as shown in Fig.4(b). The original consistency requirements are equivalently reduced to special output strategies for all the players, where the input is assigned the vertex $v_{2i-1}$ for the players $\textsf{A}_{i}, i=1, 2, 3$. This kind of graphic game has no quantum advantage for any quantum resources. It can be viewed as a relaxed graphic game satisfying partial consistency conditions in Definition 2. Here, we prove a more generalized result. Specifically, consider the following winning condition, i.e., $F(\textbf{x},\textbf{y})=1$ if $y_i=f_i(\textbf{x})$ for all $i$s; Otherwise, $F(\textbf{x},\textbf{y})=0$. Here, $f_i$ denotes some function of the input bit series $\textbf{x}$. As an example, $f_i(\textbf{x})$ can be regarded as the restriction of the product of all the assigned values in the graphic games.

{\bf Theorem 2}. There is no quantum advantage for a multipartite nonlocal game if ${\cal F}: \textbf{x} \mapsto (f_1(\textbf{x}), f_2(\textbf{x})$, $\cdots$, $f_n(\textbf{x}))$ is injective.

The proof is similar to its in Ref.[50] which is shown in Supplementary Note 3. This provides another feature to address Problem 2. Some examples are presented as follows.

{\it Example 1}. Consider $f_1, f_2, \cdots, f_n$ be any permutation in the permutation group $S_n$. Take $n=3$ as an example. All the permutations are given by
$S_3=\{(1),(1,2), (1,3),(2,3),(1,2,3),(1,3,2)\}$, where $(1)$ denotes the identity operator, $(i,j)$ denotes the permutation of $i\to j$ and $j\to i$, and $(i,j,k)$ denotes the permutation of $i\to j$, $j\to k$ and $k\to i$. Now, define the expected output of three parties as $g(i,j,k)$ for the corresponding inputs $i,j,k$ where $g\in S_3$. For the games with $(f_1,f_2,f_3)=(i,j,k)$ are previous games of guessing the neighbor' input. $^{50}$ From Theorem 2, there is no quantum advantage for this kind of games over classical resources.

{\it Example 2}. Consider $f_1, f_2, \cdots, f_n$ be any injective mapping going beyond the permutation group $S_n$, where $\{0,1\}^n$ is input space while $\mathbb{R}^n$ is output space. Take $n=3$ as an example. Consider the following mappings: ${\cal F}_1: (x_1,x_2,x_3)\mapsto (x_2+x_3,x_1+x_3,x_1+x_2)$ and ${\cal F}_2: (x_1,x_2,x_3)\mapsto (2^{x_1}-2^{x_2+x_3},2^{x_2}-2^{x_1+x_3},
2^{x_3}-2^{x_1+x_2})$. Both mappings are injective. If the mappings are used as the winning conditions of players, Theorem 2 implies no quantum advantages for these games.

{\it Example 3}. Define
$f_i=\prod_{j=1}^{k_i}s_{i_j}$, for $i=1, 2, \cdots, n$. Assume that $s_{i_1}, s_{i_2}, \cdots{},s_{i_{k_i}}$ denote special assigns on the subgarph shared by the player $\textsf{A}_j$. If $s_i\in\{\pm1\}$, the new game falls into graphic game proved in Theorem 1 when $f_i$s are injective mappings. There is no quantum advantage from Theorem 1 or Theorem 2. Interestingly, similar result holds for $s_i\in \mathbb{R}$ going beyond the games stated in Theorem 1.

\section*{Discussion}

Theorem 1 provides a result for separating quantum correlations from classical correlations in terms of multi-source networks. The consistency conditions in Definition 2 characterize the "incompatible measurements" derived from quantum mechanics. These conditions should be carefully designed for special goals in applications. Unfortunately, the graphic game cannot solve generalized XOR game which is an extension of CHSH game. $^{1,3,31}$ Another interesting problem is how to define meaningful consistency conditions such as assignments on edges going beyond Definition 2 for different goals including verifying general cyclic networks. $^{42,45}$

Generally, we presented a unified way to construct multipartite nonlocal games using graph representations of entanglement-based quantum networks. The graphic games with quantum advantage are distinguished from others for general graphs. The new model is useful for witnessing general quantum networks consisting of EPR states. This can be regarded as a new feature of multi-source networks going beyond nonlinear Bell-type inequalities and semiquantum game. Our results are interesting in Bell theory, quantum Internet, theoretical computer science, and distributive computations.

\section*{Data availability}

There is no data for the theoretical result.

\section*{Acknowledgements}

This work was supported by the National Natural Science Foundation of China (Nos.61772437,61702427), Sichuan Youth Science and Technique Foundation (No.2017JQ0048), and Fundamental Research Funds for the Central Universities (No.2018GF07).

\section*{SUPPLEMENTARY NOTE 1: THE PROOF OF THE SUFFICIENT CONDITION OF THEOREM 1}

We present the following Lemma first.

{\bf Lemma 1}. For any graph representable nonlocal game with $m=1$, assume that $p(x_i)$s have the same distribution. Then, $\varpi_q>\varpi_c$ if and only if $I_1=2$.

{\it Proof of Theorem 1}. The proof of Theorem 1 is easily from Lemma 1. Actually, $\min\{I_1, I_2,\cdots, I_m\}=2$ is equivalent to $I_j=2$. So, there exist quantum advantages over all the classical strategies with shared classical resources with respect to the subgraph involved in the observer $\textsf{A}_j$. The result can be easily extended to the global graph from the fact that $\textsf{A}_1, \textsf{A}_2, \cdots, \textsf{A}_m$ have no shared subgraphs for all the inputs being "1" from the assumption. $\square$.

The proof of Lemma 1 in Supplementary Note 1 is very long and then divided into two parts. The "if part" will be proved in Supplementary Note 1 while the "only part" will be proved in Supplementary Note 2.

{\bf Case 1. General reduction of the winning probability}

The following proof is inspired by the unbalanced CHSH game \cite{LLP,HFSD}. The key is to separate the optimal classical bound from the Tsirelson's bound of the expect winning probability \cite{Cir}. Note that Tsirelson \cite{Cir} proved that the optimal nonlocal values or quantum bounds are achievable for generalized linear games using the maximally entangled EPR states \cite{EPR}. Hence, from the equivalence of linear game and Bell inequality, it is sufficient to prove $\varpi_c<\varpi_q$ with the optimal bounds $\varpi_c$ and $\varpi_q$ when $I_1=2$. For the result of the main text, it only needs to prove that there are quantum strategies for all players such that their joint winning probability is larger than classical upper bound of $\varpi_c$, where it is very hard to find the achievable optimal quantum bound $\varpi_q$.

For input question series $\textbf{x}=x_1x_2\cdots{}x_n$, denote the product of assignments on the common vertices $V^{x_i}\cap{}V^{x_j}$ with respect to two players $\textsf{A}_i$ and $\textsf{A}_j$ as:
\begin{align*}
\zeta^{\vec{y}_{ij}}_{x_i,x_j}
=\prod_{\forall v\in{}V^{x_i}\cap {}V^{x_j}}y_{i}(v),
\tag{S1}
\\
\zeta^{\vec{y}_{ij}}_{x_i,x_j}
=\prod_{\forall v\in{}V^{x_i}\cap {}V^{x_j}}y_{j}(v),
\tag{S2}
\end{align*}
where the outputs of the player $\textsf{A}_i$ with respect to the player $\textsf{A}_j$ are respectively defined by $\vec{y}_{ij}=y_i(v_{j_1})y_i(v_{j_2}) \cdots y_i(v_{j_s})$ for $v_{j_1}, v_{j_2}, \cdots, v_{j_s} \in V^{x_i}$, which depend on the input $x_i$. The consistency condition (c) of Definition 2 of the main text is equivalent to the following equations:
\begin{align*}
\zeta^{\vec{y}_{ij}}_{x_i,x_j}
=\zeta^{\vec{y}_{ji}}_{x_j,x_i}, \forall m<i<j\leq n.
\tag{S3}
\end{align*}
Hence, the valuation function  $F(\textbf{x},\textbf{y})$ can be rewritten into
\begin{align*}
F(\textbf{x},\textbf{y})
=\prod_{i,j}\delta(\zeta^{\vec{y}_{ij}}_{x_i,x_j}
=\zeta^{\vec{y}_{ji}}_{x_i,x_j})
\tag{S4}
\end{align*}
where $\textbf{y}
=\vec{y}_1\vec{y}_2\cdots \vec{y}_n$ denote the outputs of all players, $\vec{y}_i=\vec{y}_{i\,1}\vec{y}_{i\,2}\cdots \vec{y}_{i\,n}$ denote the outputs of the player $\textsf{A}_{i}$, and $\delta(\cdot)$ denotes the characteristic function, i.e., $\delta(x)=1$ if $x$ is true and $0$ for other cases.

Now, denote all the shared resources as $\rho$, which may be a global multipartite entanglement, entangled subsystems such as EPR states, or shared classical states (hidden state model) \cite{Cir}. In what follows, we assume that all the quantum resources are EPR states \cite{EPR}. For given inputs $\textbf{x}$, denote the measurement of the player $\textsf{A}_i$ as $M_{x_i}^{\vec{y}_i}$, $i=1, 2, \cdots, n$. We have $M_{x_i}^{\vec{y}_i}=\otimes_{j=1}^n
M_{x_i}^{\vec{y}_{ij}}$, where $M_{x_i}^{\vec{y}_{ij}}$ denotes the measurement of the player $\textsf{A}_i$ on the subsystem correlated with one subsystem shared by the player $\textsf{A}_{j}$. The joint probability for the outputs $\textbf{y}$ conditional on the inputs $\textbf{x}$ is given by $P(\textbf{y}|\textbf{x})=
\langle\otimes_{i=1}^nM_{x_i}^{\vec{y}_i}\rangle
={\rm Tr}(\otimes_{i=1}^nM_{x_i}^{\vec{y}_i}\rho)$. So, any strategy with the inputs $\textbf{x}$, which satisfy the consistency condition shown in Eq.(S3), can imply the average winning probability as follows:
\begin{align*}
P_{\textbf{x}}=&
\sum_{\textbf{y}}
F(\textbf{x},\textbf{y})
P(\textbf{y}|\textbf{x})
\\
=&\sum_{\textbf{y}}
F(\textbf{x},\textbf{y})
\langle (\otimes_{i=1}^nM_{x_i}^{\vec{y}_i})
\rangle
\tag{S5}
\\
=&\sum_{\textbf{y}}
 \prod_{i<j}\delta(
 \zeta^{\vec{y}_i}_{x_i,x_j}
  =\zeta^{\vec{y}_j}_{x_i,x_j})
  \langle \otimes_{i=1}^nM_{x_i}^{\vec{y}_i}
  \rangle
\tag{S6}
\\
=&\sum_{\textbf{y}}
 \prod_{i<j}\langle \delta(\zeta^{\vec{y}_i}_{x_i,x_j}
  =\zeta^{\vec{y}_j}_{x_i,x_j})
  (M_{x_i}^{\vec{y}_{ij}}\otimes M_{x_j}^{\vec{y}_{ji}})\rangle
\tag{S7}
\\
=&\prod_{i<j}\sum_{\vec{y}_{ij}, \vec{y}_{ji} }\langle \delta(\zeta^{\vec{y}_{ij}}_{x_i,x_j}
=\zeta^{\vec{y}_{ji}}_{x_i,x_j})
(M_{x_i}^{\vec{y}_{ij}}\otimes M_{x_j}^{\vec{y}_{ji}})\rangle
\tag{S8}
\end{align*}
Eq.(S5) follows from the definition of the conditional probability $P(\textbf{y}|\textbf{x})$. Eq.(S6) is obtained by using Eq.(S4). In Eq.(S7), the expect operation $\langle M_{x_i}^{\vec{y}_{ij}}\otimes M_{x_j}^{\vec{y}_{ji}}\rangle$ is performed on the shared system of the players $\textsf{A}_i$ and $\textsf{A}_j$. Moreover, it has taken use of the equalities: $M_{x_j}^{\vec{y}_{ji}}
M_{x_j}^{\vec{y}_{ji}}
=M_{x_j}^{\hat{y}_{ji}}$ for the projection operators $M_{x_j}^{\vec{y}_{ji}}$, $i, j=1, \cdots, n$. Eq.(S8) is from the swapping of two operations by using the orthogonal conditions:  $M_{x_i}^{\vec{y}_{ij}}M_{x_i}^{\vec{y'}_{ij}}
={\bf 0}$ with $\vec{y}_{ij}\not=\vec{y'}_{ij}$ for any $x_i$, $i, j$, where $M_{x_i}^{\vec{y}_{ij}}$ and $M_{x_i}^{\vec{y'}_{ij}}$ are projection operators of the subsystem owned by the player $\textsf{A}_i$.

Note that $\delta(\zeta^{\vec{y}_{ij}}_{x_i,x_j}
=\zeta^{\vec{y}_{ji}}_{x_i,x_j})
 =\frac{1}{2}(1+\zeta^{\vec{y}_{ij}}_{x_i,x_j}
 \zeta^{\vec{y}_{ji}}_{x_i,x_j})$ for any $i, j=1, 2, \cdots, n$. It follows from Eq.(S8) that
\begin{align*}
P_{\textbf{x}}=& \prod_{i<j}\frac{1}{2}\langle(
\sum_{\hat{y}_{ij},\hat{y}_{ji}}
 M_{x_i}^{\hat{y}_{ij}}\otimes M_{x_j}^{\hat{y}_{ji}}
\\
&+\sum_{\hat{y}_{ij},\hat{y}_{ji}}
 \zeta^{\hat{y}_{ij}}_{x_i,x_j}
  \zeta^{\hat{y}_{ji}}_{x_i,x_j}
  M_{x_i}^{\hat{y}_{ij}}\otimes M_{x_j}^{\hat{y}_{ji}})\rangle
\\
=& \prod_{i<j}\frac{1}{2}\langle (\mathbb{I}
+\hat{M}_{x_i,x_j}
\otimes\hat{M}_{x_j,x_i})\rangle
\tag{S9}
\end{align*}
where $\hat{M}_{x_i,x_j}=\sum_{\vec{y}_{ij}}
\zeta^{\vec{y}_{ij}}_{x_i,x_j}
M_{x_i}^{\vec{y}_{ij}}$ denotes the successful measurement of the player $\textsf{A}_i$ on the system shared with the player $\textsf{A}_j$. $\hat{M}_{x_j,x_i}=\sum_{\vec{y}_{ji}}
\zeta^{\vec{y}_{ji}}_{x_j,x_i}
M_{x_i}^{\vec{y}_{ji}}$ denotes the successful measurement of the player $\textsf{A}_j$ on the system shared with the player $\textsf{A}_i$, for each $x_i, x_j, i,j$. In Eq.(S9), it has used the completeness of measurement operators: $\{M_{x_i}^{\vec{y}_{ij}}, \forall \vec{y}_{ij}\}$ and $\{M_{x_j}^{\vec{y}_{ji}}, \forall \vec{y}_{ji}\}$ for each $i, j$, i.e., $\sum_{\vec{y}_{ji}} M_{x_j}^{\vec{y}_{ji}}=\mathbb{I}$ and $\sum_{\vec{y}_{ij}} M_{x_i}^{\vec{y}_{ij}}=\mathbb{I}$ with the identity operator $\mathbb{I}$. The total expect winning probability is given by:
\begin{align*}
\sum_{\textbf{x}}p(\textbf{x})
P_{\textbf{x}}
 =\sum_{\textbf{x}}
 p(\textbf{x}) \prod_{i<j}\frac{1}{2}\langle (\mathbb{I}
 +\hat{M}_{x_i,x_j}
 \hat{M}_{x_j,x_i})\rangle
\tag{S10}
\end{align*}
where the tensor operation $\otimes$ is omitted for convenience because the measurement operators are clearly implemented on different local systems identified by their subindexes.

And then, Eq.(S10) is rewritten into
\begin{align*}
\sum_{\textbf{x}}
p(\textbf{x})
P_{\textbf{x}}
 =&\sum_{\textbf{x}}
 p(\textbf{x})\langle \prod_{i=2}^n\frac{1}{2}(\mathbb{I}
+\hat{M}_{x_1,x_i}\hat{M}_{x_i,x_1})
\\
&\times\prod_{j=2}^{n-1}
\prod_{k=3}^n\frac{1}{2}(\mathbb{I}
+\hat{M}_{x_j,x_k}\hat{M}_{x_k,x_j})
\rangle
\\
= &\sum_{\textbf{x}}
p(\textbf{x}) \prod_{i=2}^n\frac{1}{2}\langle \mathbb{I}
+\hat{M}_{x_1,x_i}
\hat{M}_{x_i,x_1}\rangle
\\
&\times\prod_{j=2}^{n-1}
\prod_{k=3}^n\frac{1}{2}\langle \mathbb{I}
+\hat{M}_{x_j,x_k}\hat{M}_{x_k,x_j}\rangle
\tag{S11}
\\
=&\sum_{\textbf{x}}p_*
p(\textbf{x}) \prod_{l=1}^k\prod_{j=s_l}^{n_l}
\frac{1}{2}\langle \mathbb{I}
+\hat{M}_{x_{i_l},x_j}\hat{M}_{x_j,x_{i_l}}\rangle
\tag{S12}
\end{align*}
Eq.(S12) holds for classical hidden resources by simple evaluations under the no-signalling principle, where goal is to evaluate the maximal winning probability over all the resources. For a quantum model, it is sufficient to take use of EPR states as shared resources, i.e., each pair of players $\textsf{A}_i, \textsf{A}_j$ can share some EPR states. From the assumption of the consistency condition, it is sufficient to consider the outputs of each pair of players who have common subgraphs. For other pairs, the consistency condition can be easily satisfied by setting their outputs be $1$ for all the vertices. Hence, it is reasonable to denote $p_*$ in Eq.(S12) as the success probability of all pairs without common subgraphs. It means that all the players can win the game with a constant success probability regards of their input distributions and shared resources. Here, we assume that for each $j=1, 2, \cdots, m$, all the pairs of players $\textsf{A}_{i_j}$ and $\textsf{A}_t$ have common vertices for $t=s_j, s_j+1, \cdots, n_j$, where $s_j$ and $n_j$ satisfy $1<s_1<n_1<s_1<n_2<\cdots<s_m<n_m<n$.

The following proof is divided into two cases: One is that there is no more than two players sharing vertices, and the other is for general case that there are more than two players sharing vertices shown in Supplementary Note 2.

{\bf Case 2. Proof of the sufficient part of Lemma 1}

Due to the requirements of the consistency conditions (a) and (b) in Definition 2 of the main text for the first player $\textsf{A}_1$ ($m=1$ in Lemma 1 in Supplementary Note 1), there exist four common vertex sets $V^{x_1x_j}$ for each pair of players $\textsf{A}_{1}$ and $\textsf{A}_j$ satisfying that
\begin{align*}
\hat{M}_{x_{1},x_j}&
=(-1)^{x_{1}}\hat{M}_{x_{1},1-x_j},
\tag{S13}
\\
\hat{M}_{x_j,x_1}&=
\hat{M}_{x_j,1-x_1},
\tag{S14}
\end{align*}
for binary inputs $x_1, x_j$, where  Eq.(S14) follows from the non-signaling condition.

Now, assume that there is no more than two players sharing vertices, i.e., $I_1=2$. From Eqs.(S12)-(S14), the total winning probability can be rewritten into:
\begin{align*}
\sum_{\textbf{x}}p(\textbf{x})
P_{\textbf{x}}
=& p_*[p\prod_{i=s_1}^{n_1}
\frac{1}{2}\langle \mathbb{I}
+p\hat{M}_{x_1=0,x_i}
\hat{M}_{x_i=0,x_1}
\\
&+q\hat{M}_{x_1=0,x_i}
\hat{M}_{x_i=1,x_1}\rangle
\\
&+q
\prod_{i=s_1}^{n_1}\frac{1}{2}\langle \mathbb{I}
+p\hat{M}_{x_1=1,x_i}
\hat{M}_{x_i=0,x_1}
\\
&-q\hat{M}_{x_1=1,x_i}
\hat{M}_{x_i=1,x_1}\rangle]
\\
&
\times \prod_{i=2}^{k}\prod_{j=s_i+1}^{n_i}
\frac{1}{2}\langle \mathbb{I}
+p\hat{M}_{x_i=0,x_j}
\hat{M}_{x_j=0,x_i}
\\
&+q\hat{M}_{x_i=0,x_j}
\hat{M}_{x_j=1,x_i}
+p\hat{M}_{x_i=1,x_j}
\hat{M}_{x_j=0,x_i}
\\
&
-q\hat{M}_{x_i=1,x_j}
\hat{M}_{x_j=1,x_i}\rangle]
\tag{S15}
\end{align*}
In Eq.(S15), for the simplicity we assume $i_j=j$ for $j=1, 2, \cdots, k$, $s_1=k+1$ and $s_j=n_{j-1}+1$ for $j=2, 3, \cdots, k$. Moreover, Eq.(S15) follows from the independence assumption of input distributions, where $\{p, q=1-p\}$ denotes the input distribution of each player. $\mathbb{I}+p\hat{M}_{x_i=0,x_j=0}
\hat{M}_{x_j=0,x_i=0}
+q\hat{M}_{x_i=0,x_j=1}
\hat{M}_{x_j=1,x_i=0}$ and $\mathbb{I}+p\hat{M}_{x_i=1,x_j=0}
\hat{M}_{x_j=0,x_i=1}-q
\hat{M}_{x_i=1,x_j=1}
\hat{M}_{x_j=1,x_i=1}$ are positive semidefinite operators that can be regarded as quantum observable.

To complete the proof, three subcases will be considered for following the main idea, please see Supplementary Figure 1.

\begin{figure}
\begin{center}
\includegraphics{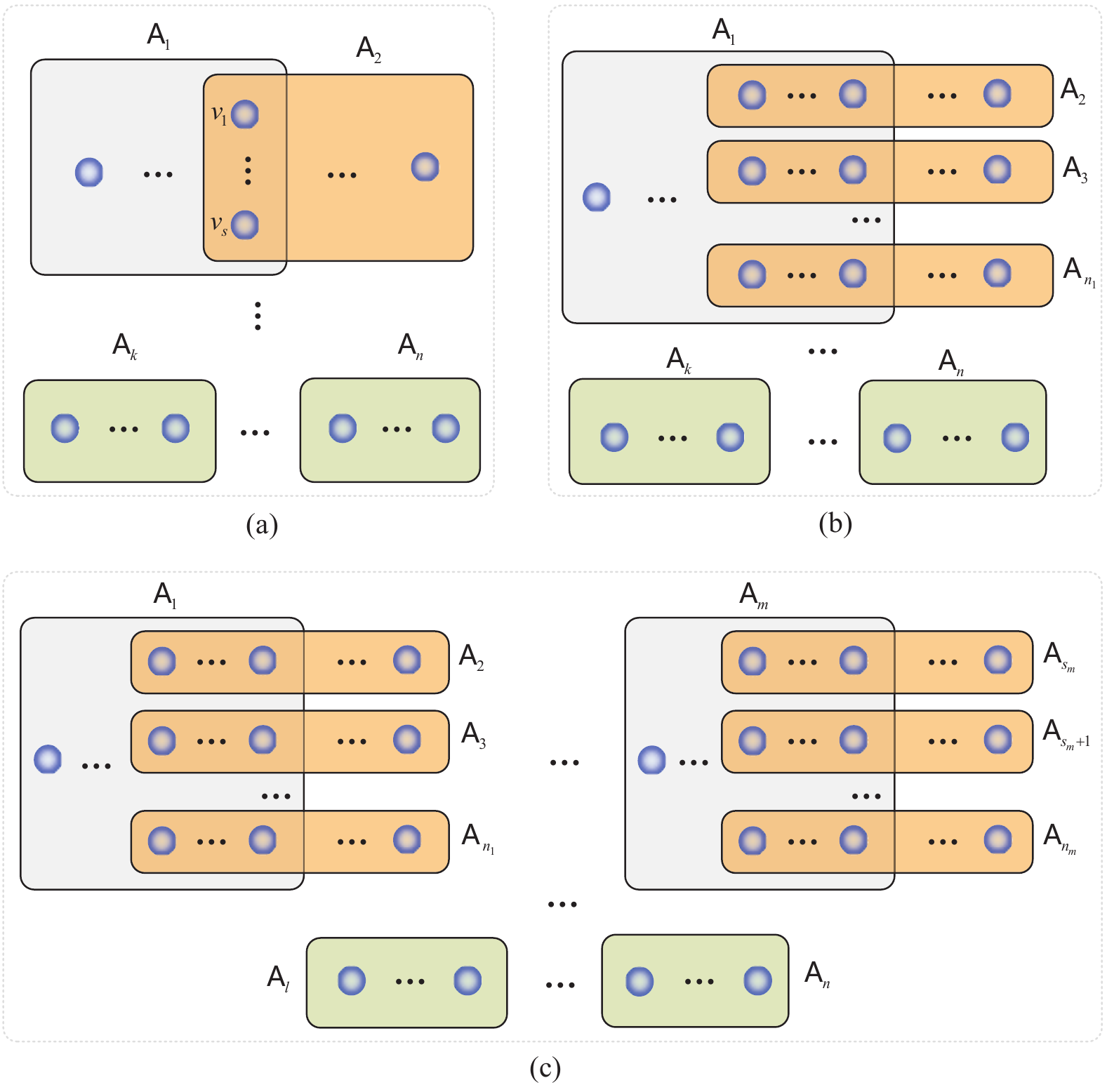}
\end{center}
\caption{(Color online) Schematic diagram of graphic games in Case one. Each player owns some vertices in a fixed graph that is known to all the players. (a) The players $\textsf{A}_1$ and $\textsf{A}_2$ share a subgraph with vertices $v_1, \cdots, v_s$ while all the other players do not share subgraph. (b) Each pair of two players $\textsf{A}_1$ and $\textsf{A}_i$ share a subgraph for $i=2, 3, \cdots, k$ while all the other players do not share subgraph. Here, any two players $\textsf{A}_i$ and $\textsf{A}_j$ share some subgraph for $i, j=2, 3, \cdots, k$. (c) Each group of players $\textsf{A}_{i_1}, \textsf{A}_{i_2}, \cdots, \textsf{A}_{i_j}$ share some subgraph with the player $\textsf{A}_{i}$ for $i=2, 3, \cdots, m$ while all the other players have no common subgraph.}
\end{figure}

{\bf Subcase 1}. $m=1$ and $s_1=n_1=2$ as shown in Supplementary Figure 1(a).

{\bf Lemma 2}. For an unbalanced CHSH game with the input probability distribution $\{p_{00},p_{01},p_{10},p_{11}\}$, the quantum advantage exists if the following conditions satisfy
\begin{itemize}
\item[(i)] $(\frac{1}{p_{10}}-\frac{1}{p_{11}})^2-(\frac{1}{p_{00}}+\frac{1}{p_{01}})^2<0$ if $\min\{p_{10},p_{11}\}\leq \min\{p_{00},p_{01}\}$;
\item[(ii)] $(\frac{1}{p_{00}}-\frac{1}{p_{01}})^2-(\frac{1}{p_{10}}+\frac{1}{p_{11}})^2<0$ if $\min\{p_{10},p_{11}\}>\min\{p_{00},p_{01}\}$.
\end{itemize}

The proof can be easily followed from its stated in Ref.\cite{LLP,CHSH}.

Hence, when $m=1$ and $n_1=2$ from Eq.(S12) it follows that
\begin{align*}
\sum_{\textbf{x}}p(\textbf{x})
P_{\textbf{x}}=
&\frac{p_*}{2}
+\frac{p_*}{2}p^2\langle
\hat{M}_{0,2}\hat{M}_{2,0}\rangle
+pq\langle\hat{M}_{0,2}
\hat{M}_{2,1}\rangle
\\
&+pq\langle\hat{M}_{1,2}\hat{M}_{2,0}\rangle
-q^2\langle\hat{M}_{1,2}\hat{M}_{2,1}\rangle,
\tag{S16}
\end{align*}
where $\hat{M}_{0,2}:=\hat{M}_{x_1=0,x_2}$, $\hat{M}_{1,2}:=\hat{M}_{x_1=1,x_2}$, $\hat{M}_{2,0}:=\hat{M}_{x_2=0,x_1}$ and $\hat{M}_{2,1}:=\hat{M}_{x_2=1,x_1}$.

From Lemma 2 in Supplementary Note 1, we can easily obtain that the quantum advantage exists for graphic games with the same input probability distribution of all players.

{\bf Subcase 2}. $m=1$ and $n_m>2$ as shown in Supplementary Figure 1(b).

In this case, suppose $s_1=2$ for convenience. Consider classical resources, i.e., $\langle\hat{M}_{x_1=0,i}\rangle$, $\langle\hat{M}_{x_1=1,i}\rangle$,
$\langle \hat{M}_{x_i=0,1}\rangle$, $\langle\hat{M}_{x_i=1,1}\rangle
\in\{\pm1\}$, from Eq.(S15) it follows that
\begin{align*}
\varpi_c:&=\max_{\lambda, \{M_{x_i}^{\hat{y}_i}\}, \forall i}\sum_{\textbf{x}}
p(\textbf{x})
 P_{\textbf{x}}
\\
&=p_*p_0+p_*(1-p_0)p_0^{n_1-1},
\tag{S17}
\end{align*}
where $p_0$ is defined by $p_0:=\max\{p,q\}$.

When entangled resources are applied for all players, it is easy to get  $(p\hat{M}_{x_1=0,i}\hat{M}_{x_i=0,1}
+q\hat{M}_{x_1=0,i}\hat{M}_{x_i=1,1})^2$
$+(p\hat{M}_{x_1=1,i}\hat{M}_{x_i=0,1}
-q\hat{M}_{x_1=1,i}\hat{M}_{x_i=1,1})^2
=2-4pq=:c^2\geq 1$ for any distributions $\{p, q\}$. Define $p\hat{M}_{x_1=0,i}\hat{M}_{x_i=0,1}
+q\hat{M}_{x_1=0,i}\hat{M}_{x_i=1,1}
=c\cos\theta_i \mathbb{I}$ and $p\hat{M}_{x_1=1,i}\hat{M}_{x_i=0,1}
-q\hat{M}_{x_1=1,i}\hat{M}_{x_i=1,1}
 =c\sin\theta_i\mathbb{I}$ w.r.s the eigenvalue decomposition of the observable. It follows from Eq.(S12) that
\begin{align*}
\varpi_q:=&\max_{\rho, \forall \{M_{x_j}^{\hat{y}_j}\}}
\sum_{\textbf{x}}
p(\textbf{x}) P_{\textbf{x}}
\\
=
&\max_{\forall \theta_j}[  \frac{p_*p}{2^{{n_1}-1}}
\prod_{i=2}^{n_1}(1+c\cos\theta_i)
+\frac{p_*q}{2^{{n_1}-1}}
\prod_{i=2}^{n_1}(1+c\sin\theta_i)]
\\
=&\frac{p_*}{2^{{n_1}-1}}
[1+\max_{\forall \theta_j}\sum_{i=i}^{{n_1}-1}
\sum_{S_i}\prod_{\theta_j\in S_i}\gamma_j
\\
&\times (p\prod_{\theta_j\in S_i}\cos\theta_j+q\prod_{\theta_j\in S_i}\sin\theta_j)]
\tag{S18}
\end{align*}
where the summation $\sum_{S_i}$ is over all the possible subsets $S_i$. Here, $S_i$ denotes the subset of $\{\theta_2, \theta_3, \cdots, \theta_{n_1}\}$ with $i$ angles $\theta_j$s.

{\bf Lemma 3} \cite{Luo}. For any $\theta_1, \theta_2, \cdots, \theta_s\in [0, \pi/2]$ and integer $s\geq 2$, the following inequality holds
 \begin{align*}
(\prod_{i=1}^s\sin\theta_i)^{\frac{1}{s}}\leq \sin(\frac{1}{s}\sum_{i=1}^s\theta_i),
\\
(\prod_{i=1}^s\cos\theta_i)^{\frac{1}{s}}\leq \cos(\frac{1}{s}\sum_{i=1}^s\theta_i),
\tag{S19}
\end{align*}
where the equality holds if and only if $\theta_1=\theta_2=\cdots =\theta_s$.

By using Lemma 3 in Supplementary Note 1, Eq.(S18) can be rewritten into
\begin{align*}
\varpi_q
  = &\frac{p_*}{2^{{n_1}-1}}
  [1+\max_{\forall \theta_j}\sum_{i=1}^{{n_1}-1}
  \sum_{S_i}\prod_{\theta_j\in S_i}
  c
\\
&\times (p\cos^i\Theta_i+q\sin^i\Theta_i)]
\tag{S20}
\\
=&\frac{p_*}{2^{{n_1}-1}}[1+\max_{\theta}
\sum_{i=1}^{{n_1}-1}\sum_{S_i}
\prod_{\theta_j\in S_i}c
\\
&\times (p\cos^i\theta+q\sin^i\theta)]
\tag{S21}
\\
=&\frac{p_*}{2^{{n_1}-1}}\max_{\theta}[
p\prod_{i=2}^{n_1}(1+c\cos\theta)
\\
&+q\prod_{i=2}^{n_1}(1+c\sin\theta)]
\tag{S22}
\end{align*}
where $\Theta_i=\frac{1}{i}\sum_{\theta_j\in S_i}\theta_j$. In Eq.(S21), the maximum achieves when the equality in inequality (S20) holds, i.e., $\theta_2=\theta_3=\cdots=\theta_{n_1}:=\theta$ from Lemma 3 in Supplementary Note 1.

Consider $p\geq 1/2$ for the simplicity. It follows from Eq.(S22) that
\begin{align*}
\varpi_q=&\frac{p_*}{2^{n_1-1}}
\max_\theta[p
 (1+c\cos\theta)^{n_1-1}
+q(1+c\sin\theta)^{n_1-1}]
\\
\geq &\frac{p_*}{2^{n_1-1}}\max_\theta[ p2^{n_1-1}(1+\frac{1}{2}(n_1-1)
(c\cos\theta-1))
\\
&+q2^{n_1-1}p^{n_1-1}
(1+(n_1-1)
\\
&\times (\frac{1}{2p}+
\frac{c}{2p}\sin\theta-1))]
\tag{S23}
\\
=&pp_*+qp_*p^{n_1-1}
+2^{n_1-2}(n_1-1)
\max_\theta[pc\cos\theta
\\
&+qp^{n_1-2}c \sin\theta+qp^{n_1-2}-2qp^{n_1-1}-p]
\\
=&p_*p+p_*qp^{n_1-1}
+p_*2(n_1-1)[c\sqrt{p^2+q^2p^{2n_1-4}}
\\
&+qp^{n_1-2}-2qp^{n_1-1}-p]
\\
>&p_*p+p_*qp^{n_1-1}
 \tag{S24} \\
=&\varpi_c
\end{align*}
when $p,q$ satisfy the following inequality:
\begin{align*}
c\sqrt{p^2+q^2p^{2n_1-4}}
+qp^{n_1-2}-2qp^{n_1-1}-p>0.
\tag{S25}
\end{align*}
Here, inequality (S23) follows from the inequality: $(1\pm x)^s\geq 1\pm sx$ for any $x\in[0,1]$ and $s\geq1$. Inequality (S24) follows from the inequality: $\max_\theta[pc\cos\theta
+qp^{n_1-2}c\sin\theta]
=c\sqrt{p^2+q^2p^{2n_1-4}}$,  where $\cos\theta=p/\sqrt{p^2
+q^2p^{2n_1-4}}$. Actually, we can prove that the inequality (S25) holds for $p\geq1/2$ and $s\geq 3$ as follows:
\begin{align*}
\Delta:=&(c-2qp^{n_1-1}-p)^2
 \\
 =&(2-4pq)(p^2+q^2p^{2n_1-4})
 -q^2p^{2n_1-4}(2p-1)^2
\\
&-2qp^{n_1-2}(2p-1)-p^2
\\
 =&q^2p^{2n_1-4}+p^2-4qp^3
 -2qp^{n_1-2}(2p-1)
\\
 =&p(p+4p^3+(2p^{-3}+4p^{-1})x+(p^{-5}+p^{-3})x^2
\\
&-2p^{-4}x^2-6p^{-2}x-4p^2)
\tag{S26}
\\
:=&f(x)
\tag{S27}
\end{align*}
In Eq.(S26), $x=p^{n_1}$. Note that $f'(x)=2(p^{-4}+p^{-2}-2p^{-3})x+2p^{-2}+4-6p^{-1}\geq0$ for any $x\in [0,1]$ because of $p^{-4}+p^{-2}-2p^{-3}\geq0$ and $2p^{-2}+4-6p^{-1}\leq 0$ for $p\in (1/2,1]$. It follows from Eq.(S26) that
\begin{align*}
\min_xf(x)=&f(p^3)
\\
=&p(2-4p-2p^2+5p^3)
\\
\geq&\frac{1}{16}>0
\end{align*}
when $x=p^3$, i.e., $n_1=3$.

Similarly, it is easy to prove $\varpi_q>\varpi_c$ when $p<1/2$. Hence, from Eqs.(S24) and (S27) we get
\begin{align*}
\varpi_q>\varpi_c.
\tag{S28}
\end{align*}

{\bf Subcase 2}. $m\geq 2$ as shown in Supplementary Figure 1(c).

This case can be easily followed by using the subcases 1 and 2, where no  consistency condition is required for $\textsf{A}_1, \textsf{A}_2, \cdots, \textsf{A}_{m}$.

\section*{SUPPLEMENTARY NOTE 2. PROOF OF NECESSARY CONDITION OF LEMMA 1}

In this note, we prove the necessary condition of Lemma 1 shown in Supplementary Note 1. Specifically, no quantum advantage exists when there are more than two players who share vertices. The proof is based on the reductions given Eqs.(S1)-(S12) shown in Supplementary Note 1. To complete the proof, consider two cases as follows: one is for the case that $l$ players $\textsf{A}_1, \textsf{A}_2, \cdots, \textsf{A}_l$ have common vertices shown in Supplementary Figure 2(a), and the other is for general case shown in Supplementary Figure 2(b).

\begin{figure}
\begin{center}
\includegraphics{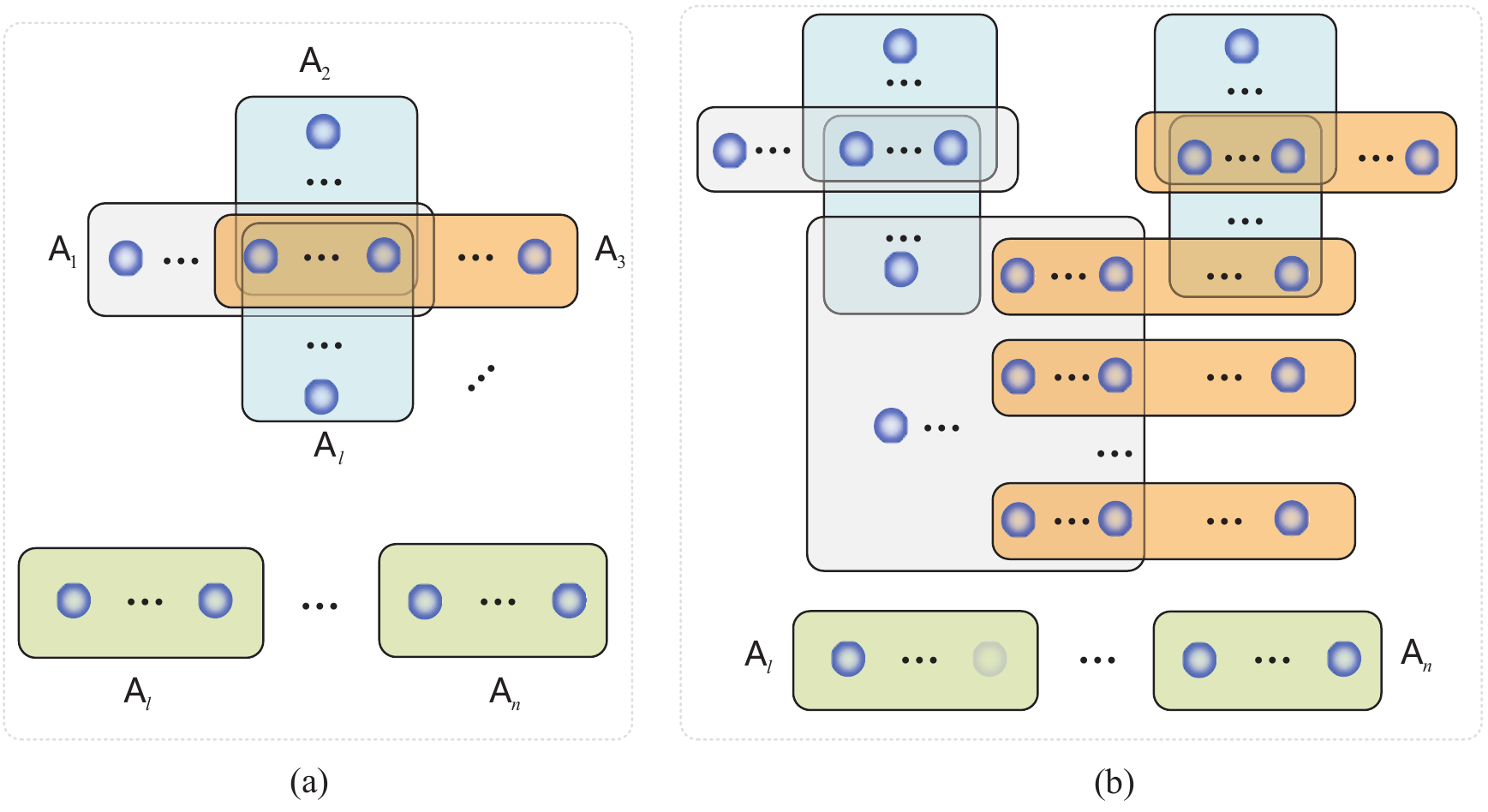}
\end{center}
\caption{(Color online) Schematic diagram of graphic games in subcase two. Each player owns some vertices in a fixed graph which is known for all the players. (a) The players $\textsf{A}_1, \textsf{A}_2, \cdots, ,\textsf{A}_l$ share the vertices $v_1, v_2, \cdots, v_s$ while all the other players do not share vertices. (b) General case. There are more than two players who share vertices. Some pair players have common vertices while some players do not share vertices.}
\end{figure}

{\bf Case 1}.  The players $\textsf{A}_1, \textsf{A}_2, \cdots, \textsf{A}_l$ have some common vertices as shown in Supplementary Figure 2(a).

In this case, we firstly assume that $l=3$ for convenience. For the classical resources, i.e., $\langle \hat{M}_{x_1=0,i}\rangle,\langle \hat{M}_{x_1=1,i}\rangle,\langle \hat{M}_{x_i=0,1}\rangle, \langle \hat{M}_{x_i=1,1}\rangle\in\{\pm1\}$, from Eq.(S12) we have
\begin{align*}
\varpi_c=&\max_{\lambda, \{M_{x_i}^{\hat{y}_i}\}, \forall i}
\sum_{\textbf{x}}
p(\textbf{x})P_{\textbf{x}}
\\
=&\max_{\lambda, \{M_{x_i}^{\hat{y}_i}\}, \forall i}(\sum_{\textbf{x}}p_*
p(\textbf{x}) [\frac{1}{2}\langle\mathbb{I}
+\hat{M}_{x_{1},2}
\hat{M}_{x_2,1}\rangle]
\\
&\times [\frac{1}{2}\langle\mathbb{I}
+\hat{M}_{x_{1},3}\hat{M}_{x_3,1}\rangle]
\times [\frac{1}{2}\langle\mathbb{I}
+\hat{M}_{x_{2},3}\hat{M}_{x_3,2}\rangle])
\tag{S29}
\end{align*}
where $p_*$ is the success probability of all the pairs without the consistency requirements.

Note that all players $\textsf{A}_1, \cdots, \textsf{A}_l$ share the same vertex set. Denote  $A_0:=\hat{M}_{x_{1}=0,2}
=\hat{M}_{x_{1}=0,3}$, $A_1:=\hat{M}_{x_{1}=1,2}
=\hat{M}_{x_{1}=1,3}$, $B_0:=\hat{M}_{x_{2}=0,1}
=\hat{M}_{x_{2}=0,3}$, $B_1:=\hat{M}_{x_{2}=1,1}
=\hat{M}_{x_{2}=1,3}$,
$C_0:=\hat{M}_{x_{3}=0,1}
=\hat{M}_{x_{3}=0,2}$, $C_1:=\hat{M}_{x_{3}=1,1}
=\hat{M}_{x_{3}=1,2}$. From Eq.(S30) it follows that
\begin{align*}
\varpi_c=&\frac{p_*}{4}\max_{\lambda, \{M_{x_i}^{\hat{y}_i}\}, \forall i}[p^3(1+\langle\,A_0B_0\rangle+\langle\,B_0C_0\rangle
+\langle\,A_0C_0\rangle)
\\
&+p^2q(1+\langle\,A_0B_0\rangle+\langle\,B_0C_1\rangle
+\langle\,A_0C_1\rangle)
\\
&+p^2q(1+\langle\,A_0B_1\rangle
+\langle\,B_1C_0\rangle+\langle\,A_0C_0\rangle)
\\
&+pq^2(1+\langle\,A_0B_1\rangle
+\langle\,B_1C_1\rangle+\langle\,A_0C_1\rangle)
\\
&+p^2q(1+\langle\,A_1B_0\rangle
+\langle\,B_0C_0\rangle+\langle\,A_1C_0\rangle)
\\
&+q^3(1-\langle\,A_1B_1\rangle
+\langle\,B_1C_1\rangle-\langle\,A_1C_1\rangle)]
\tag{S30}
\\
=&
\left\{
\begin{array}{ll}
\displaystyle
p_*(1-2p+3p^2-p^3) &\mbox{ for } 0<p\leq \frac{1}{2};
\\
p_*p(1+p-p^2) & \mbox{ for } \frac{1}{2}\leq p < 1.
\end{array}
\right.
\tag{S31}
\end{align*}
 Eq.(S31) can be proved as follows. The simulation of $\varpi_c$ is shown in Supplementary Figure 3. From the simple assumptions (deterministic strategies of all classical players) of $A_i,B_j,C_k\in\{\pm 1\}$, using Eq.(S30), it follows that
\begin{align*}
\varpi_c=\frac{p_*}{4}\max_{\lambda, \{M_{x_i}^{\hat{y}_i}\}, \forall i}[\alpha_0p^3+\alpha_1p^2q
+\alpha_2pq^2+\alpha_3q^2],
\tag{S32}
\end{align*}
where $\alpha_i$s are defined by:
\begin{align*}
\alpha_0&=1+\langle\,A_0B_0\rangle
+\langle\,B_0C_0\rangle
+\langle\,A_0C_0\rangle
\\
&\leq 4,
\\
\alpha_1&=3+\langle\,A_0B_0\rangle
+\langle\,B_0C_1\rangle
+\langle\,A_0C_1\rangle
+\langle\,A_0B_1\rangle
\\
&+\langle\,B_1C_0\rangle
+\langle\,A_0C_0\rangle
+\langle\,A_1B_0\rangle
+\langle\,B_0C_0\rangle
+\langle\,A_1C_0\rangle
\\
&\leq 12,
\\
\alpha_3&=1+\langle\,A_0B_1\rangle
+\langle\,B_1C_1\rangle
+\langle\,A_0C_1\rangle
\\
&\leq 4,
\\
\alpha_3&=1-\langle\,A_1B_1\rangle
+\langle\,B_1C_1\rangle
-\langle\,A_1C_1\rangle
\\
&\leq 4.
\end{align*}
When $p>1/2$, i.e., $p>q$, we have  $p^3,p^2q>pq^2,q^3$, and $\max \alpha_0\geq\alpha_3$, $\alpha_1>\alpha_2$. Note that $\varpi_c$ is a linear combination of four vertices $\{p^3, p^2q, pq^2, q^3\}$ for each $p, q$. The maximum of $\varpi_c$ is achievable by optimizing $\alpha_0p^3+\alpha_1p^2q$. It easily follows that $\langle\,A_0B_0\rangle
=\langle\,B_0C_0\rangle
=\langle\,A_0C_0\rangle
=\langle\,B_0C_1\rangle
=\langle\,A_0C_1\rangle
=\langle\,A_0B_1\rangle
=\langle\,B_1C_0\rangle
=\langle\,A_1B_0\rangle
=\langle\,A_1C_0\rangle
=1$. Moreover, $A_i,B_i, C_j$ are binary output observable. Similar result holds for $p<1/2$.

In what follows, it needs to prove that the quantum bound $\varpi_q$ derived from quantum measurements on any quantum resources satisfying $\varpi_q=\varpi_c$. It will be completed by three steps. Step one is used to prove the result for three players while the other is used for the case of more than three players.

\begin{figure}
\begin{center}
\resizebox{240pt}{200pt}{\includegraphics{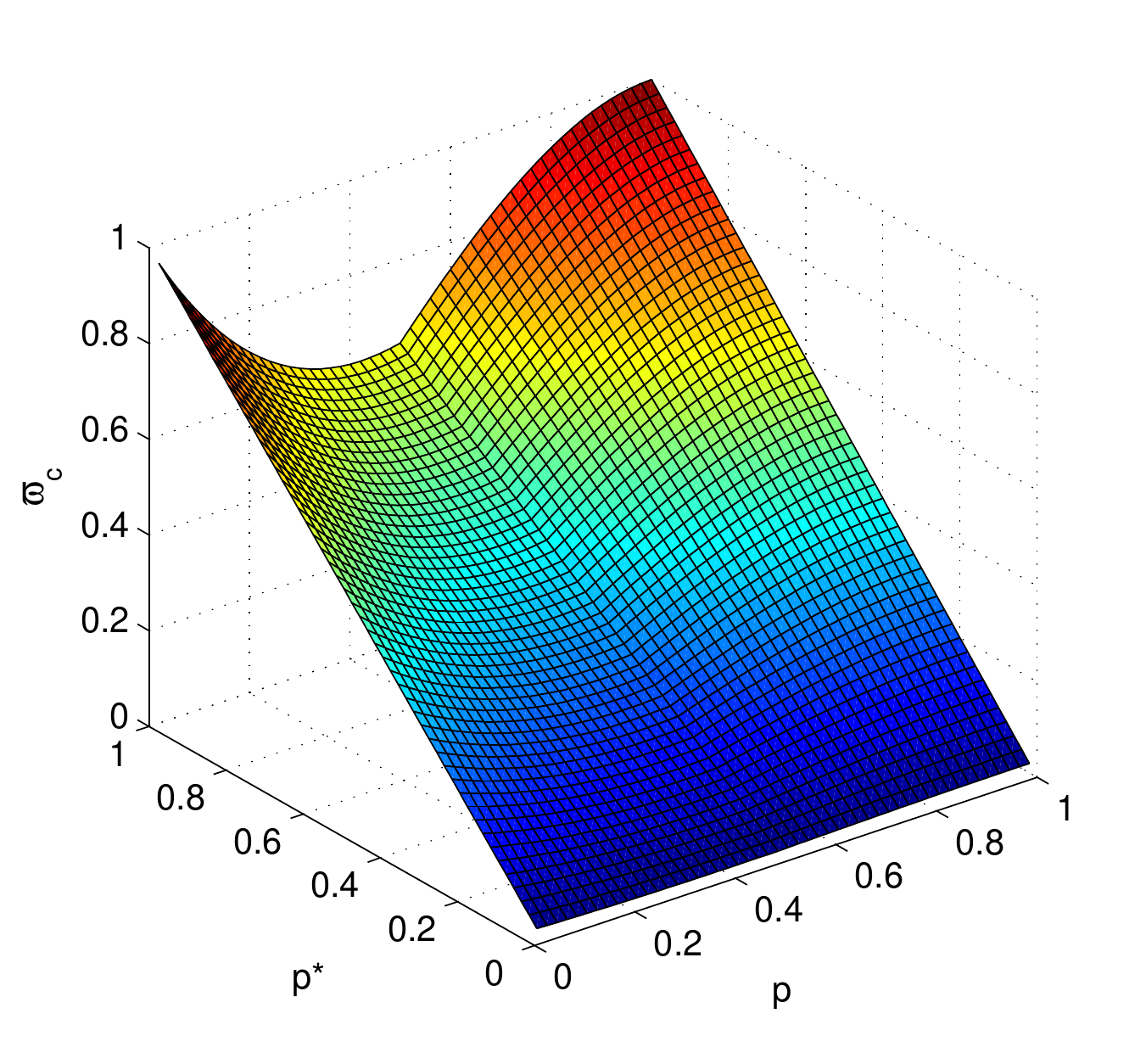}}
\end{center}
\caption{(Color online) The optimal average winning probability of classical players.}
\end{figure}

{\bf S1}. $l=3$.

Assume that the entangled resource  is $\rho$ for all the players. Note that $\hat{M}_{x_i,j}=\hat{M}_{x_i,s}$ for any $i$, $i, s \leq k$. Denote $A_{x_1}:=\hat{M}_{x_{1},2}, A_{x_1}':=\hat{M}_{x_{1},3}$, $B_{x_2}:=\hat{M}_{x_{2},1},
B_{x_2}':=\hat{M}_{x_{2},3}$, $C_{x_3}:=\hat{M}_{x_{3},1},
C_{x_3}':=\hat{M}_{x_{3},2}$ as quantum observable performed on different shared subsystems. From Eq.(S12) it follows that \begin{align*}
\varpi_q=&\max_{\rho, \{M_{x_i}^{\hat{y}_i}\}, \forall i}
\sum_{\textbf{x}}
p(\textbf{x})P_{\textbf{x}}
\\
=&\frac{1}{8}\max_{\rho, \{M_{x_i}^{\hat{y}_i}\}, \forall i}(\sum_{\textbf{x}}
p_*p(\textbf{x}) [\langle\mathbb{I}
+A_{x_{1}}B_{x_2}\rangle]
\\
&\times [\langle\mathbb{I}
+A'_{x_{1}}C_{x_3}\rangle]
\times [\langle\mathbb{I}
+B'_{x_{2}}C'_{x_3}\rangle])
\tag{S33}
\end{align*}
From the consistency conditions, for the inputs $x_1x_2x_3=101$ or $110$, the winning probability is zero since $A_{1}=B_0$, $A'_{1}=-C_1$, $B_{0}=B_0'=C_1=C_1'$, and $A_1=A'_1$ (on the common vertices of three players) ($A_{1}=-B_1$, $A'_{1}=C_0$, $A_1=A_1'$, and $B_{1}=B_1'=C_0=C_0'$) have no solution simultaneously. From consistency conditions $\hat{M}_{x_1,j}=(-1)^{x_1}
\hat{M}_{x_j,1}$ and Eq.(S33) we get that
\begin{align*}
\varpi_q
=&\frac{1}{8}p_*\max_{\rho, \{M_{x_i}^{\hat{y}_i}\}, \forall i} [p^3\langle\mathbb{I}+A_{0}B_{0}\rangle
\langle\mathbb{I}
+A'_{0}C_{0}\rangle
\\
&\times \langle\mathbb{I}
+B'_{0}C'_{0}\rangle
\\
&+
p^2q\langle\mathbb{I}+A_{0}B_{0}\rangle
\langle\mathbb{I}
+A'_{0}C_{1}\rangle
\langle\mathbb{I}
+B'_{0}C'_{1}\rangle
\\
&+p^2q\langle\mathbb{I}+A_{0}B_{1}\rangle
\langle\mathbb{I}
+A'_{0}C_{0}\rangle
\langle\mathbb{I}
+B'_{1}C'_{0}\rangle
\\
&+pq^2\langle\mathbb{I}
+A_{0}B_{1}\rangle]
\langle\mathbb{I}
+A'_{0}C_{1}\rangle
\langle\mathbb{I}
+B'_{1}C'_{1}\rangle
\\
&+p^2q\langle\mathbb{I}+A_{1}B_{0}\rangle]
\langle\mathbb{I}
+A'_{1}C_{0}\rangle
\langle\mathbb{I}
+B'_{0}C'_{0}\rangle
\\
&+q^3\langle\mathbb{I}-A_{1}B_{1}\rangle
\langle\mathbb{I}
-A'_{1}C_{1}\rangle
\langle\mathbb{I}
+B'_{1}C'_{1}\rangle
]
\\
\leq& \frac{1}{4}p_*\max_{\rho, \{M_{x_i}^{\hat{y}_i}\}, \forall i} [p^3\langle(\mathbb{I}+A_{0}B_{0})
(\mathbb{I}+A'_{0}C_{0})\rangle
\\
&+p^2q\langle(\mathbb{I}+A_{0}B_{0})
(\mathbb{I}+A'_{0}C_{1})\rangle
\\
&+p^2q\langle(\mathbb{I}+A_{0}B_{1})
(\mathbb{I}+A'_{0}C_{0})\rangle
\\
&+pq^2\langle(\mathbb{I}+A_{0}B_{1})
(\mathbb{I}+A'_{0}C_{1})\rangle
\\
&+p^2q\langle(\mathbb{I}+A_{1}B_{0})(\mathbb{I}
+A'_{1}C_{0})\rangle
\\
&+q^3\langle(\mathbb{I}-A_{1}B_{1})(\mathbb{I}
-A'_{1}C_{1})\rangle]
\tag{S34}
\\
=&\frac{1}{4}p_*\max_{\rho, \{M_{x_i}^{\hat{y}_i}\}, \forall i}[
p+p^2q+q^3+\langle\,CHSH_{AB}\rangle
\\
&+\langle\,CHSH_{AC}\rangle
+\langle{\cal L}_{ABC}\rangle]
\tag{S35}
\end{align*}
where three operators $CHSH_{AB}, CHSH_{AC}$ and  ${\cal L}_{ABC}$ are defined by:
\begin{align*}
CHSH_{AB}
=&p^2A_0B_0+pqA_0B_1+p^2qA_1B_0-q^3A_1B_1,
\\
CHSH_{AC}=&p^2A_0C_0+pqA_0C_1+p^2qA_1C_0-q^3A_1C_1,
\\
{\cal L}_{ABC}=&p^3A_0B_0C_0+p^2qA_0B_1C_0
+p^2qA_0B_0C_1
\\
&+pq^2A_0B_1C_1
+p^2qA_1B_0C_0+q^3A_1B_1C_1
\end{align*}
Here, inequality (S34) has used the inequalities: $\langle\mathbb{I}+B_iC_j\rangle\leq 2$. In Eq.(S35), we have taken use of the equalities: $A_i=A_i'$ from the consistency assumption.

Since $\|B_i\|, \|C_i\|\leq 1$, from Eq.(S35) it follows that
\begin{align}
{\cal L}_{ABC}
&\leq
\left\{
\begin{array}{ll}
{\cal L}_{AB}:=
&p^2A_0B_0+pqA_0B_1
\\
&+p^2qA_1B_0+q^3A_1B_1
\\
{\cal L}_{AC}:=
&p^2A_0C_0+pqA_0C_1
\\
&+p^2qA_1C_0+q^3A_1C_1
\end{array}
\right.
\tag{S36}
\end{align}
where all the coefficients are positive, and the maximum is  achievable by assuming $A_i, B_i\geq0$.

When $p\geq q$, $CHSH_{AB}+\frac{1}{2}{\cal L}_{AB}
=\frac{1}{2}(3p^2A_0B_0+3pqA_0B_1
+p^2qA_1B_0-3q^3A_1B_1)$ is a generalized CHSH operator. From Lemma 2 in Supplementary Note 1, the Tsirelson's bound (or quantum maximum in terms of entangled resources) is no larger than the classical maximum for $p\geq q$. It implies that no quantum advantage exists for graphic game. Similarly, it easily follows that the Tsirelson's bound of $CHSH_{AC}+\frac{1}{2}{\cal L}_{AC}$ is no larger than the classical maximum for $p\geq q$. Hence, from Eqs.(S30) and (S35), we have proved that $\varpi_q\leq \varpi_c$ for any $p, q$ with $p\geq q$.

When $p_0\leq p_1$, note that $CHSH_{AB}+\frac{1}{2}{\cal L}_{AB}
=\frac{1}{2}(3p^2A_0B_0+3pqA_0B_1
+3q^3A_1B_1-p^2qA_1B_0)$ that is a generalized CHSH operator. From Lemma 2 in Supplementary Note 1, no quantum advantage exists for graphic game with $p\leq q$. Furthermore, the Tsirelson's bound of $CHSH_{AC}+\frac{1}{2}{\cal L}_{AC}$ is no larger than its classical bound for $p\leq q$ in terms of graphic game. From Eqs.(S30) and (S35), we have proved $\varpi_q\leq \varpi_c$ for any $p, q$ with $p\leq q$. Finally, we have shown that $\varpi_q\leq \varpi_c$ for any input distributions $p, q$.

{\bf S2}. $l>3$ as shown in Supplementary Figure 2(a).

Since all the players $\textsf{A}_1, \cdots, \textsf{A}_l$ share the same vertex set, it follows that $\hat{M}_{x_{i}=0,j}=\hat{M}_{x_{i}=0,s}$ for any $i$ and $1\leq i\not=s\leq l$. From Eq.(S12) we get
\begin{align*}
\sum_{\textbf{x}}p(\textbf{x})
P_{\textbf{x}}=
&\sum_{\textbf{x}}
p(\textbf{x})P_{\textbf{x}}
\\
=&p_*\sum_{\textbf{x}}
p_*p(\textbf{x}) \prod_{1\leq i <j\leq l}(\frac{1}{2}\langle\mathbb{I}
+\hat{M}_{x_{i},j}\hat{M}_{x_j,i}\rangle)
\tag{S37}
\\
=&p_*\sum_{\textbf{x}}
p(\textbf{x}) \Delta_1\times \Delta_2\times \Delta_{3},
\tag{S38}
\end{align*}
where $\Delta_1$ is a partial summation involving the first three players ($\textsf{A}_1, \textsf{A}_2, \textsf{A}_3$), $\Delta_2$ is a partial summation involving other $l-3$ players of $\textsf{A}_4, \textsf{A}_5, \cdots, \textsf{A}_k$,  and $\Delta_3$ is a partial summation for other terms. These partial summations $\Delta_i$ are respectively defined as:
\begin{align*}
\Delta_1=& \prod_{1\leq i <j\leq 3}(\frac{1}{2}\langle\mathbb{I}
+\hat{M}_{x_{i},j}\hat{M}_{x_j,i}\rangle),
\tag{S39}
\\
\Delta_2=& \prod_{4\leq i <j\leq{}l}(\frac{1}{2}\langle\mathbb{I}
+\hat{M}_{x_{i},j}\hat{M}_{x_j,i}\rangle),
\tag{S40}
\\
\Delta_3=& \prod_{1\leq i <j\leq{}l, (i,j)\not \in {\cal I}_1\cup{\cal I}_2}(\frac{1}{2}\langle\mathbb{I}
+\hat{M}_{x_{i},j}\hat{M}_{x_j,i}\rangle)
\tag{S41}
\end{align*}
where ${\cal I}_1$ and ${\cal I}_2$ denote the respective domain of $\Delta_1$ and $\Delta_2$.

From the consistency assumption, the only nonzero winning probability for all the inputs with $x_1=1$ are $x_1x_2\cdots x_k=10\cdots 0, 11\cdots 1$. In this case, for $x_1=0$, it follows that $\Delta_2, \Delta_3\leq 1$ because of $\frac{1}{2}\langle\mathbb{I}
+\hat{M}_{x_{i},j}\hat{M}_{x_j,i}
\rangle\leq1$ for all $i,j$. Moreover, for $x_1=1$ it follows that $\Delta_2\leq 1$ because of $\frac{1}{2}\langle\mathbb{I}
+\hat{M}_{x_{i},j}\hat{M}_{x_j,i}
\rangle\leq1$ for all $i,j\geq 4$, and $\Delta_3\leq \prod_{i=4}^l
(\frac{1}{2}\langle\mathbb{I}
-\hat{M}_{x_{1},i}\hat{M}_{x_i,1}\rangle)$ since $\frac{1}{2}\langle\mathbb{I}
+\hat{M}_{x_{j},i}\hat{M}_{x_i,j}\rangle
\leq 1$ for $j=2\geq 2$ and all $i$s. Hence, when $p\geq q$, from Eqs.(S38)-(S41) we obtain that
\begin{align*}
\varpi_c:=&\max_{\lambda, \{M_{x_i}^{\hat{y}_i}\}, \forall i}\sum_{\textbf{x}}
p(\textbf{x})P_{\textbf{x}}
\\
=&p_*\max_{\lambda, \{M_{x_i}^{\hat{y}_i}\}, \forall i}(p\sum_{x_2,\cdots,x_l}
p(x_2)\cdots{}p({x_k})\Delta_1(x_1=0)
\\
&+
pq^{l-1}\Delta_1(x_1=1,x_2=x_3=0))
\tag{S42}
\\
=&p_*(p+p^{l-1}q)
\tag{S43}
\\
=&p_*(p+p^{l-1}-p^{l}).
\tag{S44}
\end{align*}
Eq.(S42) follows from the assignments of $\hat{M}_{x_{i},j}=1$ for all $i, j$. Eq.(S43) follows from the normalization condition of the distributions $p(x_1), p(x_2), \cdots, p(x_k)$. Eq.(S44) has used the equality: $p+q=1$.

Similarly, for $q\geq p$ we can obtain
\begin{align*}
\varpi_c=&\max_{\lambda, \{M_{x_i}^{\hat{y}_i}\}, \forall i}\sum_{\textbf{x}}
p(\textbf{x})P_{\textbf{x}}
\\
=&p_*(p+(1-p)^{l}).
\tag{S45}
\end{align*}

Now, for all the inputs of $x_1x_2\cdots x_{l}$, the Tsirelson's bound of $\Delta_2$ is no larger than its classical bound $1$ with $\hat{M}_{x_{i},j}=\hat{M}_{x_j,i}=1$ for $p\geq q$ or $\hat{M}_{x_{i},j}=\hat{M}_{x_j,i}=1$ except for $\hat{M}_{x_{1},j}=-1$ when $q\geq p$. From Eq.(S38), it follows that
\begin{align*}
\varpi_q=&\max_{\rho, \forall \{M_{x_j}^{\hat{y}_j}\}}
\sum_{\textbf{x}}
p(\textbf{x})
 P_{\textbf{x}}
\\
\leq &p_*\max_{\rho, \forall \{M_{x_j}^{\hat{y}_j}\}}[
\sum_{x_2,x_3}pp(x_2)p(x_3)
\Delta_1(x_1=0)
\\
&+qp^{l-1}\Delta_1(x_1=1,x_2=x_3=0)
\\
&+q^{l}\Delta_1(x_1=x_2=x_3=1)
\prod_{i=4}^l(\frac{1}{2}\langle
\mathbb{I}
-\hat{M}_{x_{1},i}\hat{M}_{x_i,1}\rangle)]
\\
\leq &p_*\max_{\rho, \{\forall M_{x_i}^{\hat{y}_i}\}}
[\sum_{x_2,x_3}pp(x_2)p(x_3)
\Delta_1(x_1=0)
\\
&+qp^{l-1}\Delta_1(x_1=1,x_2=x_3=0)
\\
&+q^{l}\Delta_1(x_1=x_2=x_3=1)]
\tag{S46}
\\
:=&\frac{1}{4}p_*\max_{\rho, \{\forall M_{x_i}^{\hat{y}_i}\}}[
p+p^{l-1}q+q^{l}
+\langle\widehat{CHSH}_{AB}\rangle
\\
&+\langle\, \widehat{CHSH}_{AC}\rangle
+\langle\widehat{\cal L}_{ABC}\rangle],
\tag{S47}
\end{align*}
where we have taken the notions of $A_i, B_i, C_i$ defined in Eq.(S35), and three partial operators are defined as:
\begin{align*}
\widehat{CHSH}_{AB}=
&p^2A_0B_0+pqA_0B_1
+p^{l-1}qA_1B_0-q^{l}A_1B_1,
\\
\widehat{CHSH}_{AC}=& p^2A_0C_0
+pqA_0C_1+p^{l-1}qA_1C_0
-q^{l} A_1C_1,
\\
\widehat{\cal L}_{ABC}=& p^3A_0B_0C_0+p^2qA_0B_1C_0
+p^2qA_0B_0C_1\\
&
+pq^2A_0B_1C_1
+p^{l-1}qA_1B_0C_0+q^{l}A_1B_1C_1.
\end{align*}
Eq.(S46) has used the inequalities:  $\frac{1}{2}\langle\mathbb{I}
-\hat{M}_{x_{1},i}\hat{M}_{x_i,1}
\rangle\leq 1$ for all $i$. Note that
\begin{align}
\widehat{\cal L}_{ABC}
&\leq
\left\{
\begin{array}{ll}
\widehat{\cal L}_{AB}:=
&p^2A_0B_0+pqA_0B_1
\\
&+p^{l-1}qA_1B_0+q^{l}A_1B_1
\\
\widehat{\cal L}_{AC}:=
&p^2A_0C_0+pqA_0C_1
\\
&+p^{l-1}qA_1C_0+q^{l}A_1C_1
\end{array}
\right.
\tag{S48}
\end{align}
Similar to Eq.(S35), the operator bound (quantum bound) of the generalized CHSH operators  $\widehat{CHSH}_{AB}+\frac{1}{2}
\widehat{\cal L}_{AB}$ and $\widehat{CHSH}_{AC}+\frac{1}{2}
\widehat{\cal L}_{AC}$ is no larger than its classical bound. Moreover, it is easy to prove that $p_*(p+p^{l-1}-p^{l})$ and $p_*(p+(1-p)^{l})$ are the optimal classical bounds of the corresponding generalized operators defined in Eq.(S47) for $p\geq q$ and $p\leq q$, respectively. We obtain that
\begin{align*}
\varpi_q \leq &
\left\{
\begin{array}{lll}
p_*(p+p^{l-1}-p^{l}) \mbox{ for } p\geq q,
\\
p_*(p+(1-p)^{l})  \mbox{ for } p\leq q.
\end{array}
\right.
\tag{S49}
\end{align*}
Consequently, from Eq.(S44), (S45) and (S49) we proved that
\begin{align*}
\varpi_q\leq \varpi_c.
\tag{S50}
\end{align*}
It means that no quantum advantage exists for graphic game in this case.

\section*{SUPPLEMENTARY NOTE 3. THE PROOF OF THEOREM 2}

From Theorem 1 of the main text, the graphic game with proper restrictions or consistency conditions has quantum advantage. Conversely, it is not true even if for the uniform inputs. Specifically, there are different games with only one or two common vertices, which have quantum advantage or not. So, it requires further restrictions for characterizing these games without quantum advantage. One example is guessing your neighbor game \cite{GYNI}. Here, we provide a generalized result. Specifically, we consider the winning conditions, i.e., $F(\textbf{x},\textbf{y})=1$ if $a_i=f_i(\textbf{x})$ for all $i$s; Otherwise, $F(\textbf{x},\textbf{y})=0$. Here, $f_i$ denotes some function of the input $\textbf{x}$. As an example, $f_i(\textbf{x})$ can be regarded as the restriction of the product of all the assigned values in the graphic games.

{\bf Theorem 2}. There is no quantum advantage for multipartite nonlocal game if ${\cal F}: \textbf{x} \mapsto (f_1(\textbf{x}), f_2(\textbf{x}), \cdots, f_n(\textbf{x}))$ is injective.

{\bf Proof of Theorem 2}. Inspired by the quantum game \cite{GYNI}, we show that the optimal classical and quantum winning strategies are identical for any prior distribution $p$ of the inputs. Note that there is a simple classical strategy achieving a winning probability
\begin{align*}
\varpi_c\geq\max_{\textbf{x}}
(p(\textbf{x}) + p(\overline{\textbf{x}}))
 \tag{S51}
\end{align*}
where $\overline{\textbf{x}}$ denotes the negation of the input string $\textbf{x}$, $\overline{\textbf{x}}= (\overline{x}_1, \overline{x}_2, \cdots, \overline{x}_n)$ with $\overline{x}_i =x_i\oplus1$, and $\oplus$ denotes addition modulo 2. This strategy is based on the following simple observation from the assumption of ${\cal F}$.
\begin{itemize}
\item[]Let $\textbf{x}^*$ be an arbitrary string. If $\textbf{x} \not=\textbf{x}^*$, $\overline{\textbf{x}^*}$, then it follows that $(f_1(\textbf{x}^*), \cdots, f_n(\textbf{x}^*))
    \not=(f_1(\textbf{x}), \cdots,f_n(\textbf{x}))$, and $(f_1(\overline{\textbf{x}^*}), \cdots,f_n(\overline{\textbf{x}^*}))
    \not=(f_1(\textbf{x}), \cdots,f_n(\textbf{x}))$. $\qquad$  (S52)
\end{itemize}

Indeed, if this was not the case, we would have that for any $i$, either $x_i\not=y_i$ or $x_{i+1}=y_{i+1}$. But this would in turn imply that either $\textbf{x}
=\textbf{x}^*$ or $\textbf{x}
=\overline{\textbf{x}^*}$, which is contradiction with the assumption.

Let a classical strategy be specified by the string $\textbf{x}^*$, where each party sends out the bit $y_i=f_{i}(\textbf{x}^*)$ if it received the input $x^*_i$, and outputs $y_i =f_{i}(\overline{\textbf{x}^*})$ if it received $\overline{x^*_i}$. It obviously follows that $P(y_i=f_{i}(\textbf{x}^*)
|\textbf{x}^*)=1$ and $P(y_i=f_{i}(\overline{\textbf{x}^*})
|\overline{\textbf{x}^*})=1$. On the other hand, $P(y_i|\textbf{x})= 0$ for all $\textbf{x}\not=
\textbf{x}^*, \overline{\textbf{y}}$. Indeed, from the observation (S52), there exists an index $i$ such that $x_i=x^*_i$, but for which $x^*_{i+1}\not=x_{i+1}$. Since ${\cal F}$ is injective, it follows that $(f_1(\textbf{x}^*), f_2(\textbf{x}^*), \cdots, f_n(\textbf{x}^*))
\not=(f_1(\textbf{x})$,
$f_2(\textbf{x})$,
$\cdots, f_n(\textbf{x}))$, and $(f_1(\overline{\textbf{x}^*}), f_2(\overline{\textbf{x}^*}), \cdots, f_n(\overline{\textbf{x}^*}))
\not=(f_1(\textbf{x}),
f_2(\textbf{x}), \cdots, f_n(\textbf{x}))$. The winning probability of this classical strategy equals to $p(\textbf{x}^*)+ p(\overline{\textbf{y}})$, which yields Eq.(S51) if we take $\textbf{x}^*$ to be $p(\textbf{x}^*)
+p(\overline{\textbf{x}^*})
=\max_{\textbf{x}}
(p(\textbf{x}) +p(\overline{\textbf{x}}))$.

We now prove that there is no better quantum (hence classical) strategy. In the most general quantum protocol, the players share an entangled state $\rho$ and perform projective measurements on their subsystem depending on their inputs $x_i$. They output their measurement results $y_i$. Denote $M^{x_i}_{y_i}$ as the projection operator associated to the output $y_i$ for the input $x_i$. The probability that all players produce the correct output is given by
\begin{align*}
&P(y_1=f_1(\textbf{x}),\cdots,y_n =f_n(\textbf{x})|\textbf{x}) \\
=& \langle M^{x_1}_{f_1}\otimes \cdots\otimes M^{x_n}_{f_n}\rangle
\\
=&{\rm Tr}[(M^{x_1}_{f_1}\otimes \cdots\otimes M^{x_n}_{f_n})\rho],
\tag{S53}
\end{align*}
The average winning probability is
\begin{align*}
\varpi_q=\max_{\rho,\forall
\{M^{x_i}_{y_i}\}}
\sum_{\textbf{x}}
p(\textbf{x})\langle M_{\textbf{x}}\rangle
\tag{S54}
\end{align*}
where $M_{\textbf{x}}$ is given by $M_{\textbf{x}}
=M^{x_1}_{f_1}\otimes \cdots\otimes M^{x_n}_{f_n}$ for short, and $\langle \cdot \rangle$ is associated with some fixed quantum system $\rho$. Note that the operators $M_{\textbf{x}}$ satisfy the following properties
\begin{align*}
&M^2_{\textbf{x}} =
M_{\textbf{x}},
\tag{S55}
\\
&M_{\textbf{x}}
M_{\textbf{x}^*}=0 \mbox{ if } \textbf{x} \not=\textbf{x}^*,
\overline{\textbf{x}^*}.
\tag{S56}
\end{align*}
The first property follows from the fact that $M_{\textbf{x}}$s are projection operators. The second follows from the orthogonality relations: $M^{x_i}_{y_i}
M^{x_i}_{\overline{y}_i}=0$ and observation in Eq.(S52). Note that protocols involving mixed states or general measurements can all be represented in the above form by expanding the dimensionality of the initial state.

Using Eqs.(S55) and (S56), we show that $\varpi_q=\max_{\rho,\forall
\{M^{x_i}_{y_i}\}}
\sum_{\textbf{x}}
p(\textbf{x})
M_{\textbf{x}}\leq \varpi_c$, where "$\leq$" should be understood as an operator inequality, i.e., $A\leq B$ means that $\langle A\rangle\leq \langle B\rangle$ for all $\rho$. Note that we cannot assume $p(\textbf{x})+p(\overline{\textbf{x}})
= \varpi_c$, but $p(\textbf{x})+ p(\overline{\textbf{x}})\leq \varpi_c$ for all $\textbf{x}$ since the normalization condition of $\sum_{\textbf{x}}
p(\textbf{x})
=1$. In fact, we have
\begin{align*}
\Delta=&(\sqrt{\varpi_c}-
\sum_{\textbf{x}}
\alpha_{\textbf{x}}
M_{\textbf{x}})^2
+\frac{1}{2}\sum_{\textbf{x}}
(\beta_{\textbf{x}}
M_{\textbf{x}}-
\beta_{\overline{\textbf{x}}}
M_{\overline{\textbf{x}}})^2
\\
=&\varpi_c-2\sum_{\textbf{x}}
(\varpi_c-p(\overline{\textbf{x}}))
M_{\textbf{x}}
\\
&+\sum_{\textbf{x}}
\alpha_{\textbf{x}}^2
M_{\textbf{x}}
+\sum_{\textbf{x}}
\alpha_{\textbf{x}}
\alpha_{\overline{\textbf{x}}}
M_{\textbf{x}}M_{\overline{\textbf{x}}}
\\
&+\frac{1}{2}\sum_{\textbf{x}}
\frac{p(\textbf{x})
p(\overline{\textbf{x}})}{\varpi_c}
(M_{\textbf{x}}+M_{\overline{\textbf{x}}}
\\
&-M_{\textbf{x}}
M_{\overline{\textbf{x}}}
-M_{\overline{\textbf{x}}}
M_{\textbf{x}})
\tag{S57}
\end{align*}
where $\alpha_{\textbf{x}}=
\sqrt{\varpi_c}-p(\overline{\textbf{x}})
/\sqrt{\varpi_c}$ and $\beta_{\textbf{x}}
=\sqrt{p(\textbf{x})
p(\overline{\textbf{x}})/\varpi_c}$. Eq.(S57) is followed from Eq.(S55). And then, we have
\begin{align*}
\Delta=&\varpi_c-
2\sum_{\textbf{x}}(\varpi_c-
p(\overline{\textbf{x}}))
M_{\textbf{x}}+\sum_{\textbf{x}}
\frac{p^2(\overline{\textbf{x}})}{
\varpi_c}M_{\textbf{x}}
\\
&+\sum_{\textbf{x}}
\varpi_{c}M_{\textbf{x}}
-2\sum_{\textbf{x}}
p(\overline{\textbf{x}})
M_{\textbf{x}}
\\
&+\sum_{\textbf{x}}(\varpi_c-
p(\overline{\textbf{x}})
-p(\textbf{x})
+\frac{p(\textbf{x})
p(\overline{\textbf{x}})}{\varpi_c})
M_{\overline{\textbf{x}}}
M_{\textbf{x}}
\\
&+\frac{1}{2}\sum_{\textbf{x}}
\frac{p(\textbf{x})
p(\overline{\textbf{x}})}{\varpi_c}
(M_{\textbf{x}}
+M_{\overline{\textbf{x}}}
-M_{\textbf{x}}
M_{\overline{\textbf{x}}}
-M_{\overline{\textbf{x}}}
M_{\textbf{x}})
\\
=&\varpi_c-\sum_{\textbf{x}}
\varpi_cM_{\textbf{x}}
+\sum_{\textbf{x}}
(\varpi_c-p(\textbf{x})
-p(\overline{\textbf{x}}))
M_{\textbf{x}}M_{\overline{\textbf{x}}}
\\
&+\sum_{\textbf{x}}\frac{p(\overline{\textbf{x}})
(p(\textbf{x})
+p(\overline{\textbf{x}}))}{
\varpi_c}M_{\textbf{x}}
\tag{S58}
\\
\leq &\varpi_c-\sum_{\textbf{x}}
\varpi_cM_{\textbf{x}}
+\sum_{\textbf{x}}
p(\overline{\textbf{x}})
M_{\textbf{x}}
\\
&+\sum_{\textbf{x}}
(\varpi_c-p(\textbf{x})
-p(\overline{\textbf{x}}))
M_{\textbf{x}}
M_{\overline{\textbf{x}}}
\tag{S59}
\\
=&\varpi_c-\sum_{\textbf{x}}
(\varpi_c-p(\overline{\textbf{x}}))
M_{\textbf{x}}
\\
&+\sum_{\textbf{x}}
(\varpi_c-p(\textbf{x})
-p(\overline{\textbf{x}}))
M_{\textbf{x}}M_{\overline{\textbf{x}}}
\\
\leq &\varpi_c-\sum_{\textbf{x}}(
\varpi_c-p(\overline{\textbf{x}}))
M_{\textbf{x}}
\\
&+\sum_{\textbf{x}}
(\varpi_c-p(\textbf{x})
-p(\overline{\textbf{x}}))
M_{\textbf{x}}
\tag{S60}
\\
=&\varpi_c-\sum_{\textbf{x}}
p(\textbf{x})M_{\textbf{x}},
\tag{S61}
\end{align*}
In Eq.(S58), we have taken use of the commutate of projection operators, i.e., $M_{\textbf{x}}
M_{\overline{\textbf{x}}}
=M_{\overline{\textbf{x}}}
M_{\textbf{x}}$, and identity $\sum_{\textbf{x}}
\frac{p(\overline{\textbf{x}})
p(\textbf{x})}{\varpi_c}
M_{\overline{\textbf{x}}}
=\sum_{\textbf{x}}
\frac{p(\textbf{x})
p(\overline{\textbf{x}})}{\varpi_c}
M_{\textbf{x}}$. Eq.(S59) follows from the inequality : $p(\textbf{x})
+p(\overline{\textbf{x}})\leq \varpi_c$ which is derived from the inequality (S51). Eq.(S60) follows from the operator inequality of  $M_{\textbf{x}}
M_{\overline{\textbf{x}}}\leq M_{\textbf{x}}$. From Eq.(S51), it follows that
\begin{align*}
\max_{\rho,\forall\{M^{x_i}_{y_i}\}}
\sum_{\textbf{x}}
p(\textbf{x})
M_{\textbf{x}}\leq \varpi_c.
\tag{S62}
\end{align*}

The inequality of $\sum_{\textbf{x}} p(\textbf{x})P(y_1=f_1(\textbf{x}), \cdots, y_n=f_n(\textbf{x})
|\textbf{x})\leq \varpi_c$ can be interpreted as a Bell inequality whose local and quantum bound coincide. Note that any POVM measurements associated with non-commuting operators can be realized using projection measurements in an extended state space. The proof stated above is also independent of quantum resources. This explains the same bound of the classical and quantum payoffs.


\end{document}